\documentclass[twocolumn,showpacs,prc,nofootinbib]{revtex4}

\usepackage{epsfig}
\usepackage{graphicx}
\usepackage{amssymb}
\usepackage{amsmath}

\def\be{\begin{eqnarray}}
\def\ee{\end{eqnarray}}
\def\MeV{\mbox{MeV}}

\def\lsim{\lesssim}
\def\gsim{\gtrsim}
\def\la{\langle}
\def\ra{\rangle}
\newcommand{\msun}{\mbox{$M_\odot$}}

\def\del{\partial}

\newcommand{\ba}{\begin{eqnarray}}
\newcommand{\ea}{\end{eqnarray}}

\begin{document}



\title{Dense Stellar Matter with Strange Quark Matter Driven by Kaon Condensation}
\author{Kyungmin Kim$^{1,}$\footnote{kmkim82@hanyang.ac.kr}, Hyun Kyu Lee$^{1,}$\footnote{hyunkyu@hanyang.ac.kr} and Mannque Rho$^{1,2,}$\footnote{mannque.rho@cea.fr} }
\affiliation{
\medskip
  $^1$Department of Physics, Hanyang University, Seoul 133-791, Korea\\
}
\affiliation{
  $^2$Institut de Physique Th$\acute{e}$orique, Direction des Sciences de la Mati$\grave{e}$re, CEA Saclay,\\ 91191 Gif-sur-Yvette, France\\
}
\date{\today}


\begin{abstract}
 The core of neutron-star matter is supposed to be at a much higher density than the normal nuclear-matter density, for which various possibilities have been suggested, such as, for example, meson or hyperon condensation and/or deconfined quark or color-superconducting matter. In this work, we explore the implication on hadron physics of a dense compact object that has three ``phases":  nuclear matter at the outer layer, kaon condensed nuclear matter in the middle, and strange quark matter at the core. Using a drastically simplified but not unreasonable model, we develop the scenario where the different phases are smoothly connected with the kaon condensed matter playing a role of a ``doorway" to a quark core, the equation of state of which with parameters {\em restricted within the range allowed by nature} could be made compatible with the mass vs. radius constraint given by the 1.97-solar-mass object PSR J1614-2230 recently observed.
\end{abstract}

\maketitle

\section{Introduction}
The recent observation of a 1.97-solar-mass ($M_{\odot}$) neutron star, PSR J1614-2230, using Shapiro delay with Green Bank Telescope \cite{2solar} , raises a highly pertinent issue on whether the non-nuclear degrees of freedom such as meson condensations (pions, kaons, and/or hyperons), quark matter (normal or color-superconducting), and the multitude of other phases are relevant for the physics of stable compact stars. This observation, if unequivocally confirmed, will have far-reaching ramifications on sorting out the large variety of models proposed in the literature for the maximum (minimum) mass of neutron stars (black holes). A particularly poignant example is the scenario of Bethe and Brown\cite{bb}\cite{BLR-PR}, where the onset of kaon condensation inside the neutron star matter at a density $\rho\sim 3\rho_0$ -- where $\rho_0$ is the nuclear matter density -- keeps the maximum mass less than $2 M_{\odot}$, which seems to be consistent with the observations of well-measured neutron-star masses $\sim 1.5 M_{\odot}$~\cite{1.5obs}, beyond which, it is argued, the stars collapse to black holes, giving rise to an enhanced number of light-mass black holes in the universe. At a much more drastic level, there is also the issue of possible non-Newtonian gravity intervening with ``soft" equations of state that may otherwise be ruled out by the observation~\cite{non-newton}. This circle of issues was recently raised and analyzed, among others, by Lattimer and Prakash~\cite{LP-GFEST}.

Unlike hadronic matter at high temperature for which lattice QCD provides model-independent control of effective theories, cold baryonic matter at high density has remained a largely unchartered domain, more or less inaccessible by model-independent approaches. Because of the sign problem, lattice QCD cannot yet probe the relevant density and furthermore there are no known model-independent tools available for treating baryonic matter at high density. The abundant and accurate experimental data on finite nuclei allow one to determine the equation of state (EoS) for both symmetric and asymmetric systems up to the normal nuclear-matter density $\rho_0$. However, there is neither a reliable theoretical tool nor unambiguous experimental information to enable one to extrapolate the EoS beyond $\rho_0$ in a controlled way. This conundrum accounts for the widely diverging model predictions available in the literature.

In this paper, we would like to address the EoS of compact-star matter, with the observation of the PSR J1614-2230 specifically in mind, along the line that has been developed in anticipation of what is to come from the laboratories in construction for the purpose of probing cold  baryonic matter at high density such as FAIR/GSI and J-PARC. The question we are raising is as follows: Is it possible to reconcile the existence of a 2 $M_{\odot}$ neutron star with an EoS that encompasses normal nuclear matter, strange hadronic matter, and quark matter in a continuous phase diagram? In addressing this question, it is, of course, not in our capability to make a fully consistent treatment backed by rigorous arguments. As precisely set forth in the next section, what is empirically known in hadron dynamics up to the normal nuclear-matter density $\rho_0$ that enters into the EoS is interpreted in the simplest possible way and extrapolated in the given scheme to the unknown higher density regime relevant to compact stars, guided by whatever available terrestrial as well as astronomical observables. What we do is, while sticking as much as is feasible to extreme simplicity,  develop a chain of arguments which are {\it not obviously wrong} and could perhaps be quickly falsified in the laboratories. In this way, we bypass most of the complications invoking ``reasonableness." Certain caveats inherent in such an approach are pointed out in the Appendix. {Needless to say, what is badly needed is a unified framework in which the different phases involved in the process in the compact-star structure are consistently connected.}

 We should mention that there have been a large number of treatments in the literature that attempt to take into account a variety of relevant degrees of freedom, some of which overlap with what we consider here and their interactions, as reviewed in Ref. \cite{LP-review}. What may differentiate our approach from others is in the -- perhaps too extreme -- simplification of the complexity involved in the processes while preserving a certain level of consistency and that most -- if not all -- of the assumptions we make could be falsifiable.

Our starting point is the obvious fact that the neutron-star matter possesses the $n$-$p$ asymmetry in its composition, the effect of which on EoS is manifested in the ``symmetry energy," denoted hereafter $E_{sym}.$\footnote{For those who are not in the field, it should be more aptly called  ``asymmetry energy" because it is the energy that the asymmetry between the neutron and proton numbers costs. We follow the convention of the community of this field by calling it symmetry energy.} It also determines the threshold for new degrees of freedom that can emerge from weak interactions. The larger the symmetry energy, the more energy the asymmetry between neutron and proton composition costs.  By introducing new degrees of freedom into the system, one can then lower the energy of the system and relax the asymmetry between neutron and proton.  For instance, the energy of a pure neutron star matter can be lowered by introducing electrons, allowing protons to appear to compensate the charge, which, in turn, relaxes the cost of the asymmetry owing to the symmetry energy. So unless there is any mechanism that keeps the matter from  relaxing the neutron-proton asymmetry present,  the system will evolve to a matter in a more stable state by bringing in new degrees of freedom.
The change of neutron and proton fractions of the star matter would require the isospin to be violated, which is, of course, not allowed by the strong interactions.  However, this can happen via the weak interactions, where the flavor (isospin or strangeness) can be changed. We assume that the evolution of the system via weak interactions reaches an equilibrium configuration as a ground state of the matter.  Therefore, the EoS of the star matter in weak equilibrium must be strongly dependent on the symmetry energy, especially for the emergence of new degrees of freedom accompanied by the change of the neutron-proton fraction.   One of the most interesting scenarios is that kaon condensation can occur inside the neutron star as a specific example of a new degree of freedom. We focus on this in this paper although other strangeness degrees of freedom -- when treated differently as specified below -- can also be important.

In this work, we investigate the role of kaon condensation as the principal agent for strangeness in evolving to the core of compact stars that exists in the form of strange quark matter. Put more precisely, kaon condensation will play a role of a ``doorway" to quark matter in the sense precisely stated in the next section. Focusing on the symmetry energy, we consider compact stars that consist of nuclear matter (NM) in the outer-layer region and (1) kaon condensed nuclear matter (KNM) and/or (2) strange quark matter (SQM) in the inner region. We explore how the structure with  KNM and SQM in the core region affects the microscopic observables [e.g., density-dependent electron (or kaon) chemical potential] and macroscopic observables (e.g., mass and radius) of the compact stars.

The outline of this paper is as follows. First we state as precisely as possible the strategy that we take for our approach. This is the content of Sec. II. The global features of kaon condensation and SQM that appear as the degrees of freedom extraneous to nucleons is given in Sec. III.  In Sec. IV, the highly simplified model for kaon condensation that we use is described. We also discuss the microscopic properties of nucleonic matter with kaon condensation and the EoS.  The stellar structure of this system obtained by integrating the Tolman-Oppenheimer-Volkov (TOV) equation is given in Sec. V.   In Sec. VI, with the given SQM EoS, we study the stellar structure with SQM driven by kaon condensation. {The effects of QCD perturbative corrections are also discussed.}  Throughout this work, we set $\hbar = c = 1$ and assume zero temperature. Finally, the results are summarized and discussed in Sec. VII.

\section{The Strategy}\label{strategy}
The system we are interested in involves three basic interactions: gravity, weak interactions and strong interactions. We assume the first two are known, so our focus is on the third. We are assuming that solving the TOV equation with a given EoS takes care of the gravity interaction. Both weak interactions and strong interactions figure in the EoS. In considering the strong interactions, we are concerned -- among others -- with three possible phases, namely, NM, strange hadronic matter, or, more specifically, KNM and SQM. Neutron stars, of course, have matters other than strong interactions, such as the crust which must be considered on the same footing. We do not deal with these matters in this work.

The compact star we are particularly interested in consists of three layers: the outer layer of NM, the intermediary layer of kaon condensation, and the inner core of SQM. Because we do not know how to formulate the problem within the framework of QCD, let us imagine having at our disposal a theory that has a hadronic Lagrangian ${\cal L}_{hadron}$ applicable up to some density $\rho_c$, say, the chiral transition density or deconfinement density, and then above that density, an effective Lagrangian ${\cal L}_{quark}$ that contains QCD variables, that is, quarks and gluons. Apart from the asymptotic density most likely inaccessible by compact stars, we know very little of what ${\cal L}_{quark}$ encodes. It could be a color-flavor locked superconducting state, Overhauser state, dyonic crystal, etc. We simply assume, as is done by other workers, that it can be captured by the MIT bag model with or without perturbtive QCD corrections.

That leaves our principal problem in this paper: ${\cal L}_{hadron}$.

In the hadronic sector below $\rho_c$, several density scales must intervene. The first is the NM density denoted $\rho_0$ and the second is the density at which strange degrees of freedom start figuring, in our case, kaon condensation at a density denoted $\rho_t$. There is also -- in a specific model to be specified below -- a possible topological phase transition at a density denoted $\rho_{1/2}$ lying above $\rho_0$ and below or above $\rho_t$. This phase is not understood well at the moment, so it is not taken into account in the present work. Its possible ramifications in compact-star physics are discussed in the Appendix.

Here is the strategy we take in addressing these different scales. Up to the NM density, the effective hadronic Lagrangian ${\cal L}_{NN}$ that encodes nuclear physics is mostly phenomenological, with the physics involved relying on available experimental data. In fact, most of the EoS used in the literature have been determined to be consistent with the available experimental data up to the normal nuclear density $\rho_0$, -- so all of them more or less agree with each other. However, above $\rho_0$,  quite arbitrary extrapolations are involved. They are somewhat constrained by certain experiments that can probe densities slightly higher than $\rho_0$. However, the error band of the available experimental data is quite large, making it difficult to pin down the structure. The presently available density regime goes up to $\sim 4\rho_0$ with rather large error bands.\footnote{For an extensive review on the matter on which we heavily rely in our work, see \cite{bal}.} We will pick one particular phenomenological parametrization that approximately covers the range of the experimental error band up to the density ``probed" $\sim 4\rho_0$. As to the density regime beyond it, we have no guidance or control, so we will simply extrapolate it in the form we pick. Several potentially important effects that are predicted by a given field theoretic framework based on hidden local symmetry but cannot be simply implemented in the strategy we are taking are mentioned in the Appendix.

At least up to $\rho_0$, the strange degrees of freedom play no role,\footnote{This is obviated by the lattice QCD indication that the strangeness content of the nucleon is negligible.} but they must figure at some density above $\rho_0$ in the guise of kaons and hyperons. In addressing this issue, we will take a simplified effective field theory approach. Instead of treating both kaons and hyperons at the same time, which is quite involved numerically, we may integrate out either the kaons or the hyperons in the given effective Lagrangian. Now if we integrate out the kaons (as well as vector mesons), then the resulting effective Lagrangian will consist of a sum of three-flavor [i.e., $SU(3)_f$] local four-Fermi (baryon) interactions. Integrating out the kaons whose mass becomes small as density increases is, in a sense, analogous to integrating out the pions from chiral effective field theories for describing nuclear processes whose energy scale is much smaller than the pion mass.\footnote{This eminently ``non-standard" approach to low-energy nuclear physics is called ``pionless effective field theory"~\cite{pionless}.} Doing mean field with this four-Fermi effective Lagrangian is equivalent to doing mean field with the original baryon $SU(3)_f$ chiral Lagrangian in a way similar to the Walecka mean-field theory that is obtained from four-Fermi chiral Lagrangian~\cite{Ritzi}.

Alternatively, we can integrate out the hyperons from the Lagrangian. This will give us an effective Lagrangian consisting of kaons coupled to nucleons with the couplings determined by the integrating-out procedure. When treated at mean field (i.e., at tree order), this would correspond to the mean-field treatment of chiral Lagrangian first put forward by Kaplan and Nelson~\cite{kaplan-nelson} with, however, the coefficients treated accordingly. As is elaborated in Sec. \ref{kaoncon}, there is a hint from lattice QCD calculation, that owing to large cancellation between higher-order terms, the tree-order treatment survives more or less unscathed. Assuming this mechanism would render focusing on the kaon sector much simpler than looking at the hyperon sector. In this paper, we adopt this option.

To be fair, we should point out that there has been a long controversy, still unresolved, on the relevance of kaon condensation vs. hyperon condensation in compact-star structure. We have no firm argument to support but we deem it plausible that both approaches mentioned above capture essentially the same physics -- and not a clear-cut alternative -- vis-\`a-vis to the role of strangeness in the dense strange-matter sector of the star. In a more complete formulation where both degrees of freedom are treated explicitly and on the same footing, there will be sharing of effects between the two and most of the observables may not be able to disentangle one from the other. Future observations of neutrino emission from the core might give a hint to the roles of these degrees of freedom.

In sum, our key question is as follows: Within the uncertainties present both in the obesrvables and in the parameters of the model,  is it possible to develop a scenario of stellar matter consisting of NM, kaon condensed matter, and quark matter with smooth transitions from one to the other?

\section{Kaon Condensation and Strange Quark Matter}
In this section, we present a simple description of how kaon condensed matter can be linked to SQM.

In stellar matter, the chemical equilibrium in weak interaction leads to a condition for the chemical potentials of neutron, proton, electron, and muon:
\be
\mu_n - \mu_p = \mu_e = \mu_{\mu} \equiv \mu.
\ee
The density dependence of neutron-proton chemical potential difference is determined by the symmetry energy in the following form:
\be
\mu_n - \mu_p = 4(1 - 2 \frac{\rho_p}{\rho})S(\rho)\ .
\ee
Here $\rho$ is the sum of proton density and neutron density, $\rho=\rho_p +\rho_n$, and $S(\rho)$ is the symmetry-energy factor that is specified later.

\subsection{Kaon condensation}
 As is discussed below, the key ingredient in our treatment is the decrease of the effective mass denoted as $m_K^*$\footnote{In this paper,  $m^*_K$ is the kaon energy in medium because we are dealing with the s-wave kaon.} of the negatively charged kaon $K^-$ as density increases.  The $m_K^*$ is basically a function of $m_K,\rho_n, \rho_p$ owing to the kaon-nucleon interactions:
 \be
 m_K^*= \omega(m_K,\rho_n, \rho_p, ...). \label{mko}
 \ee
 The density at which a neutron can decay into a proton and $K^-$ via the weak process, $n \rightarrow p + K^-$,
\be
\mu_n - \mu_p = m_K^*,
\ee
determines the threshold of kaon condensation $\rho_t$. Above the kaon condensation, where $ m_K^*$ can be identified as the kaon chemical potential $\mu_K$, the chemical equilibrium is reached as
\be
\mu_n - \mu_p = \mu_e = \mu_{\mu} = \mu_K \equiv \mu.
\ee
where
\be
\mu_n - \mu_p&=& 4(1 - 2 \frac{\rho_p}{\rho})S(\rho) + \Theta(K)F(K,\mu), \label{betaeq0}
\ee
where $K$ stands for the kaon amplitude of kaon condensed state, i.e., $\la K\ra$,  and $F(K,\mu)$ is a nontrivial function that depends on the neutron-proton chemical potential difference which, in turn, depends on kaon-nucleon interactions. It is a highly model-dependent quantity.
The charge neutrality condition gives
\be
\rho_p =\rho_e + \rho_{\mu} +  \Theta(K)\rho_K\ .  \label{neutral}
\ee
Equations (\ref{betaeq0}), (\ref{neutral}) and (\ref{mko}) are the basic equations to be solved to calculate the EoS of KNM.

\subsection{Strange quark matter}
Strange quark matter can appear as a result of confinement-deconfinement phase transition.  This phase transition is basically a strong interaction process.  Suppose NM with no strangeness changes over to a deconfined quark phase. There is no room for strange quarks in this case because it is the strong interaction that is involved. However, in dense compact-star matter, the system evolves in weak equilibrium with strange quarks in the quark matter through weak interactions.  The nature of interaction for the phase change is, however, different if it is from a phase in which  kaons are present to start with induced by the weak interaction. Now if kaon condensation has taken place in the system, then there is already non-vanishing strangeness in the NM up to the phase boundary. In this case,  we can imagine the confinement-deconfinement phase transition taking place constrained by the weak equilibrium in the kaon condensed matter, leading to SQM. In this subsection, we develop the scenario in which kaon condensation joins smoothly a SQM.

We first consider NM (without kaon) and deconfined quark matter (QM). Suppose there is a phase boundary between NM and QM at the density $\rho_c$ (for NM) and $\rho^Q_c$ (for QM),\footnote{We are taking into consideration that there may be a density jump between the NM and the QM at the boundary, such that $\rho^Q_c\neq 3\rho_c$.} respectively, in the interior of a neutron star. The chemical equilibrium reads
\be
\mu_n = 2\mu_d + \mu_u, ~~~ \mu_p =  \mu_d + 2\mu_u. \label{chemeq}
\ee
The symmetry energies in each phase are related to the chemical potentials as
\be
\mu_n-\mu_p &=& 4 (1-2x_N) S_N(\rho_N)\\
\mu_d-\mu_u &=& 4 (1-2x_Q) S_Q(\rho_Q),
\ee
where $x_N = \rho_p/\rho_N$, $x_Q = \rho_u/\rho_Q$, and $\rho_Q = \rho_u + \rho_d$.
From the chemical equilibrium
\be
\mu_n -\mu_p &=& \mu_d-\mu_u,
\ee
using $x_Q = \rho_u/\rho_Q = (1+x_N)/3$, we get
\be
(1-2x_N) S_N(\rho_c) = \frac{1}{3}(1-2x_N) S_Q(\rho^Q_c) \label{xn}
\ee
from which we obtain a constraint on the symmetry energy at $\rho_c$,
\be
S_N(\rho_c)&=& \frac{1}{3}S_Q(\rho^Q_c),
\ee
for $x_N \neq 1/2$.
Just for an illustration, suppose we take an expression for the symmetry energy factor of the non-interacting quark gas,
\be
S_Q = \frac{3}{4}(3 \pi^2 \rho_Q /2 )^{1/3}[2^{1/3} -1],
\ee
and assume that the critical quark number density, $\rho^Q_c / 3$, is equal to the critical baryon number density, $\rho_c$;
then we will get
\be
S_N(\rho_c) = 24.6 (\frac{\rho_c}{\rho_0})^{1/3} \MeV\ . \label{snnc}
\ee
We may obtain a different constraint on the possible form of nuclear symmetry energy, if NM undergoes a phase transition into a quark phase at $\rho_c$. However, that will yield a highly nontrivial constraint, which seems not to be realized\footnote{In a recent paper by G. Pagliara and J. Schaffner-Bielich \cite{PSB}, it is argued that the $S_Q$ can be three times larger than that given by the free quark model if the quark phase is in 2SC, which is, of course, a highly correlated system, unlike free quarks.}.   In Eq.~(\ref{xn}), the simplest solution is $x_N =1/2$ irrespective of the specific form of symmetry energy. We obtain this result from a different consideration.

Now consider the case where SQM meets KNM.   As was done in the previous section, we can obtain the kaon threshold density, $\rho_t$, by the profile of the kaon chemical potential. From $\rho_t$ up to the critical density for chiral restoration $\rho_c$, the weak equilibrium condition reads
\be
\mu=4(1 - 2 \rho_p/\rho)S_N(\rho_c) + F(K,\mu,\rho_c).  \label{betaeq3}
\ee
Now introduce the Weinberg-Tomozawa term for the kaon-nucleon interactions which are introduced in Section \ref{kaoncon}. Since $F(K,\mu,\rho_c)$ can be expressed as
\be
F(K,\mu,\rho_c) = \mu \tilde{F}(K,\rho_c),
\ee
we have
\be
\mu=4(1 - 2 \rho_p/\rho)S_N(\rho_c) + \mu \tilde{F}(K,\rho_c).  \label{betaeq2}
\ee
As the kaon chemical potential -- equivalently effective mass -- $\mu$ approaches 0 at the critical density,  the solution $x=\frac{\rho_p}{\rho}=1/2$  appears naturally at the phase boundary,  provided $\tilde{F}(K,\rho_c) \neq 1$, which is assumed to be valid for the range of density we are concerned with.

The chemical equilibrium (via confinement-deconfinement) reads
\be
\mu_n -\mu_p &=& \mu_d-\mu_u,\label{chemq} \\
\mu_{K^-} &=& \mu_s - \mu_u \label{chemk}
\ee
at the phase boundary.
Note that the strange quark is required at the boundary, which implies that there should be a SQM for $\rho > \rho_c$.
Because $\mu(=\mu_K) =0$ , we have from Eqs.~(\ref{chemq}) and (\ref{chemk})
\be
\mu_u ~ = ~ \mu_d \ \ {\rm and} \ \ \mu_s ~ = ~ \mu_u,
\ee
which gives
\be
\mu_u = \mu_d = \mu_s. \label{musqm}
\ee
This is the chemical potential relation for the SQM in the masselss limit.  In this simple picture, the KNM leads naturally to a SQM.

\section{Effective Theory for Kaon Condensation}
\subsection{The effective Lagrangian}\label{kaoncon}
Following the strategy spelled out in Section \ref{strategy}, we consider kaon condensation in the simplest form of an effective chiral Lagrangian. Projecting onto the $K^-$ channel ignoring all other (pseudo-) Goldstone bosons and the nucleons with hyperons integrated out, we write the relevant effective Lagrangian in the form~\cite{kuniharu}
\be
{\cal L} = {\cal L}_{KN} + {\cal L}_{NN} \label{Leff}\ee
where
\be
{\cal L}_{KN} &=& \partial_{\mu} K^- \partial^{\mu} K^+ - m^2_K K^+ K^- \nonumber\\
&+& \frac{1}{f^2} \Sigma_{KN}(n^{\dagger} n + p^{\dagger}p) K^+ K^-\nonumber \\
&+& \frac{i}{4f^2} (n^{\dagger} n + 2 p^{\dagger}p) (K^+\partial_{0} K^- - K^- \partial_{0} K^+) \label{LKN}\\
{\cal L}_{NN} &=&
   n^{\dagger} i \partial_{0} n +  p^{\dagger} i \partial_{0} p - \frac{1}{2m}( \vec{\nabla} n^{\dagger}\cdot \vec{\nabla} n + \vec{\nabla} p^{\dagger} \cdot \vec{\nabla} p) \nonumber\\
   &-& V_{NN} \label{LNN}
\ee
Here the fourth term in Eq.~(\ref{LKN}) is the well-known Weinberg-Tomozawa (WT) term which is constrained by a low-energy theorem with $f$ identified with the pion decay constant $f_\pi$ in the {\em matter-free} space and $\Sigma_{KN}$ is the $KN$ sigma term
\be
\Sigma_{KN}\approx  \frac 12 (\bar{m}+m_s)\la N|\bar{u}u+\bar{s}s|N\ra\ \label{sigmaterm}
\ee
where $\bar{m}=(m_u+m_d)/2$ with the subscripts $u$, $d$, and $s$ standing, respectively, for up-quark, down-quark, and strange quark.
In chiral perturbation theory, the WT term corresponds to the ${\cal O}(p)$ term and the sigma term to ${\cal O}(p^2)$. In principle, there is no reason why one should stop at that order in the chiral counting and also to quadratic order in the kaon field as is done in Eq.~(\ref{LKN}). In fact there is a large literature that treats the chiral Lagrangian to high orders in addressing kaon-nuclear physics, but we shall getting into the complexity of such sophisticated approaches.

Although we are unable to make rigorous arguments, we can offer several reasons why we can take the simplest form Eq.~(\ref{LKN}) as our basis of analysis. From the point of view of renormalization group flow, what matters is the WT term with the $\Sigma$ term being irrelevant. This was shown in Ref. \cite{LRS}. There it was shown that it is the WT term coming with attraction in kaon-nuclear interactions that drives kaon condensation. That consideration does not determine the critical density. It simply indicates that kaon condensation is bound to take place at increasing density. It also indicates that the $\Lambda (1405)$ which plays a crucial role for the threshold $KN$ interactions has little role in kaon condensation transition itself. The discussion on these matters -- still a highly controversial issue -- is relegated to Appendix.

That the most important interaction for kaon-nuclear physics in the chiral Lagrangian is the WT term is exploited in the approach of Ref. \cite{weise}, where all higher chiral order terms are ignored and only the WT term is inserted as a driving term in a coupled-channel -- with $\pi\Sigma$ channel -- equation. In a similar spirit, if one approaches kaon condensation from the vector-manifestation fixed point~\cite{HY:PR} rather than from the matter-free space, one can ignore the $\Sigma$ term because the quark condensate vanishes and retain only the light vector-meson-exchange terms (i.e., the WT term)~\cite{BLPR}. The question would arise if the $\Sigma$ term were as large as thought in chiral perturbation approach. In the early days, it was thought that the strangeness content of the nucleon related to $m_s\la N|\bar{s}s|N\ra$ was substantial, say, $m_s\la N|\bar{s}s|N\ra > 300$ MeV. Thus, in ChPT, the $\Sigma$ term played the dominant mechanism for the condensation. Nuclear density played a role of eating into the kaon mass; thus, the ``large" $\Sigma$ term  figuring importantly in the nucleon structure has an essential role. Viewed from the matter-free space, it is the $\Sigma$ term that is responsible for the relatively low kaon condensation density $\rho_K\sim 3\rho_0$. This feature is being drastically changed by the recent lattice results that find that $m_s\la N|\bar{s}s|N\ra \lesssim 59$ MeV~\cite{lattice}, which is about one-sixth of the old value used in the literature~\cite{BLR-PR}. This would give a sigma term $\Sigma_{KN}\lesssim 200$ MeV.
Furthermore, as suggested when viewed from the vector manifestation fixed point of view, the sigma term may be strongly suppressed by density near the condensation point because the in-nucleon condensate $\la N|\bar{q}q|N\ra$  will be suppressed, as argued in Ref. \cite{BLPR}. It seems that the importance of the sigma term in the old treatment of kaon condensation was overstated or even unfounded. Our point of view is that the sigma term Eq.~(\ref{sigmaterm}) is still quite uncertain at high density.

Next we address the issue of limiting kaon fields to quadratic order in Eq.~(\ref{LKN}) for the EoS we are interested in. Within the framework, one might attempt to calculate a few next orders of chiral perturbation -- and this has been done in the literature --, but there is no compelling reason why a few next orders give reliable results. As it appears to be generally the case in applying chiral perturbation theory to nuclear and dense matter, it may be that going to a few higher-order terms that are calculable starting from matter-free vacuum does not capture the essential aspect of the process.  Indeed a recent pioneering lattice QCD calculation~\cite{savage} indicates that this may be the case. In this reference, lattice QCD simulation of kaon condensation on dense kaon systems containing up to 12 $K^-$'s showed that the properties of the condensate are remarkably well reproduced by leading-order chiral perturbation theory. While direct baryonic background, difficult to implement in lattice calculations, is still needed, what this surprising result indicates, rather persuasively, is that there can be substantial cancelations in higher chiral-order terms that cannot be captured by a few perturbatively calculable terms. While awaiting for further more quantitative guidance from model-independent theory and/or experiments, we take the following philosophy in handling the Lagrangian Eq.~(\ref{LKN}). Adopting it for our approach -- which is essentially a mean-field one, we simply take the WT as is in medium. This is because that term is constrained by low-energy theorems. Given the wide uncertainty with the sigma term, we will consider $\Sigma_{KN}$ as a free parameter constrained by the low value given by the lattice results $\lsim 200$ MeV: It can be as low as $\sim 130$ MeV depending on the $\pi$-N sigma term as it seems to be required for certain nuclear EoS and as high as $\sim 400$ MeV, as was taken in the old days.
\subsection{The effective Hamiltonian}
Along the strategy described in Sec. \ref{strategy} for the kaon-nucleon interaction as well as for the nucleonic interaction, we write the Hamiltonian in the form
\be
{\cal H} = {\cal H}_{KN}  + {\cal H}_{NN},
\ee
where
\be
{\cal H}_{KN} &=&  \partial_{0} K^- \partial^{0} K^+  + [m^2_K  -
\frac{n}{f^2} \Sigma_{KN}] K^+ K^-, \label{hkn}\\
{\cal H}_{NN} &=& \frac{3}{5}E_F^0 \left(\frac{\rho}{\rho_0}\right)^{2/3}
\rho + V(\rho) + \rho \left(1-2\frac{\rho_p}{\rho}\right)^2 S(\rho). \nonumber\\
\ee
Here $V(\rho)$ is the potential energy of NM as a function of density $\rho$ and $S(\rho)$ is the symmetry energy factor as a function of nuclear density. Both $S(\rho)$ and $V(\rho)$ are model dependent. There are numerous forms for them in the literature, all fine-tuned to fit available experimental data~\cite{bal}. What matters for our purpose is that they be consistent with experiments up to the density that has been probed. The data are not precise enough above NM density and hence there is a range of parameters that are consistent with the error bands given by the data. What we will do is to pick one convenient parametrization called MDI (``momentum-dependent interaction") from the review~\cite{bal}, which we refer to as ``LCK''. It has the form\footnote{The notation for the variable $x$ in Eqs.(5.1)-(5.5) of \cite{bal} is changed to $\eta$ in this paper, because $x$ is used for the proton number fraction in the text.}
\be
S(\rho,\eta) &=& F(\eta)\frac{\rho}{\rho_0}+[18.6-F(\eta)]\left(\frac{\rho}{\rho_0}\right)^{G(\eta)} \nonumber\\
&+& (2^{2/3}-1)\frac{3}{5}E_F^0\left(\frac{\rho}{\rho_0}\right)^{2/3}, \label{bals} \\
V(\rho) &=& [m_N + \frac{3}{5}E_F^0 \left(\frac{\rho}{\rho_0}\right)^{2/3}] \rho +\frac{\alpha}{2}\frac{\rho}{\rho_0} \nonumber\\
&+& \frac{\beta}{1+\gamma}\left(\frac{\rho}{\rho_0}\right)^{\gamma}, \label{balv}
\ee
where $\alpha=-298.25\MeV$, $\beta=244.99 \MeV$, and $\gamma=1.21$.
The remaining parameters, $F(\eta)$ and $G(\eta)$, for this model are summarized in Table 1. We take two values for $\eta$, i.e., $\eta=0, -1$. These two values are compatible with the band of empirical constraints up to density $\sim 4\rho_0$, with $\eta=-1$ representing stiffer EoS with the classic AKR (Akmal-Pandharipande-Ravenhal) EoS~\cite{APR} sandwiched by the two $\eta$ values.\footnote{We do not consider $\eta=1$, which gives what is called ``supersoft" symmetry energy, which by itself is not consistent with experimental constraints. See \cite{non-newton}.} We see that while consistent within the error band with experimental constraints, they can give different results at higher density in the EoS when the strangeness and quark degrees of freedom are involved.
\begin{table}[h]
\begin{center}
\begin{minipage}{.4\textwidth}
\begin{center}
\begin{tabular}{c | c | c }
\hline
\hline
Model & $F(\eta)$   & $G(\eta)$ \\
\hline
$\eta=-1$ & $3.673$ & $1.569$ \\
\hline
$\eta=0$ & $129.981$ & $1.059$ \\
\hline
\hline
\end{tabular}
\caption{The parameters for $\eta=-1$ and $\eta=0$. $F(\eta)$ is given in $\MeV$ unit and $G(\eta)$ is dimensionless quantity.}
\end{center}
\end{minipage}
\end{center}
\label{nl3param}
\end{table}

\subsection{S-wave kaon condensation }
Using the chemical potentials defined by
\be
\mu_i = \frac{\del {\cal H}}{\del \rho_i},
\ee
the chemical potentials of proton and neutron are given by
\be
\mu_n &=& \mu_n^0 - \left[ \frac{\mu}{2 f^2} + \frac{\Sigma_{KN}}{f^2} \right] K^2, \\
\mu_p &=& \mu_p^0 - \left[ \frac{\mu}{f^2} + \frac{\Sigma_{KN}}{f^2} \right] K^2,
\ee
where
\be
\mu_n^0 - \mu_p^0 = 4 (1 - 2\frac{\rho_p}{\rho}) S(\rho).
\ee
The amplitude of kaon condensation, $K$, and kaon chemical potential, $\mu_K$, are defined by the ansatz
\be
K^{\pm} = K e^{\pm i \mu t}.
\ee
The kaon condensation condition for $K \neq 0$ is obtained by extremizing the classical action,
\be
m_K^2 - \mu^2 = \mu \frac{\rho_n + 2\rho_p}{2f^2} + \frac{\rho}{f^2}\Sigma_{KN}, \label{kaoncond}
\ee
which can be solved to get $\mu$ or equivalently $m_K^*$ in Eq.~(\ref{mko}).

\begin{figure}[t!]
\begin{center}
\includegraphics[width=8.6cm]{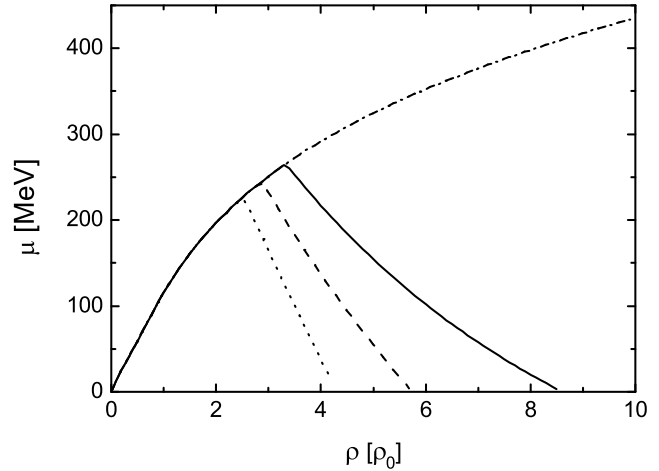}
\caption{The density dependence of the chemical potential for LCK with $\eta=-1$. The solid, dashed, and dotted lines correspond, respectively, to $\Sigma_{KN}=200,300, 400$ MeV.}
\label{chemsbal1}
\end{center}
\end{figure}

\begin{figure}[t!]
\begin{center}
\includegraphics[width=8.6cm]{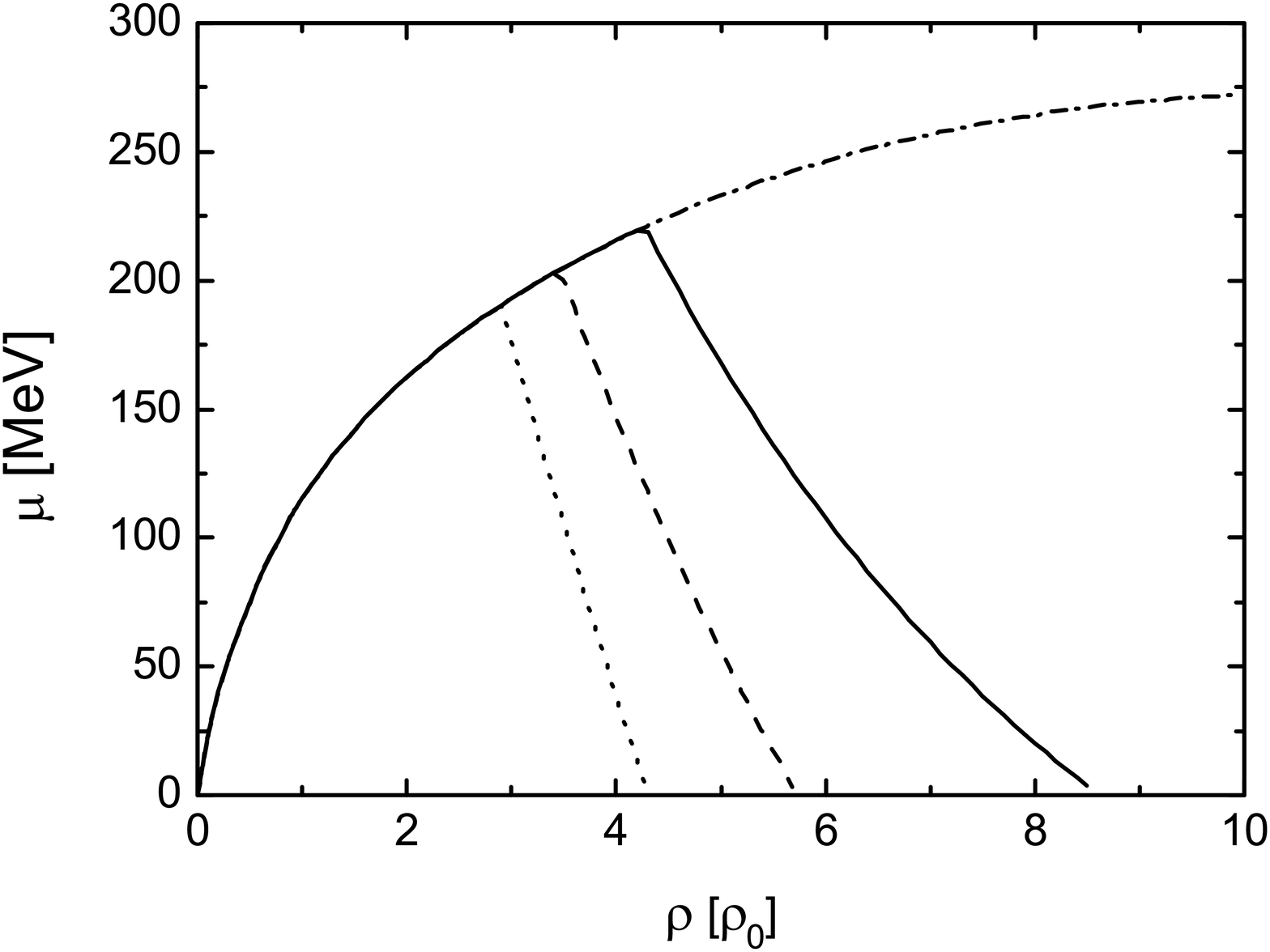}
\caption{The density dependence of the chemical potential for LCK with $\eta=0$. The solid, dashed, and dotted lines correspond, respectively, to $\Sigma_{KN}=200,300, 400$ MeV, as in Fig.~\ref{chemsbal1}.}
\label{chemsbal2}
\end{center}
\end{figure}

\begin{figure}[t!]
\begin{center}
\includegraphics[width=8.6cm]{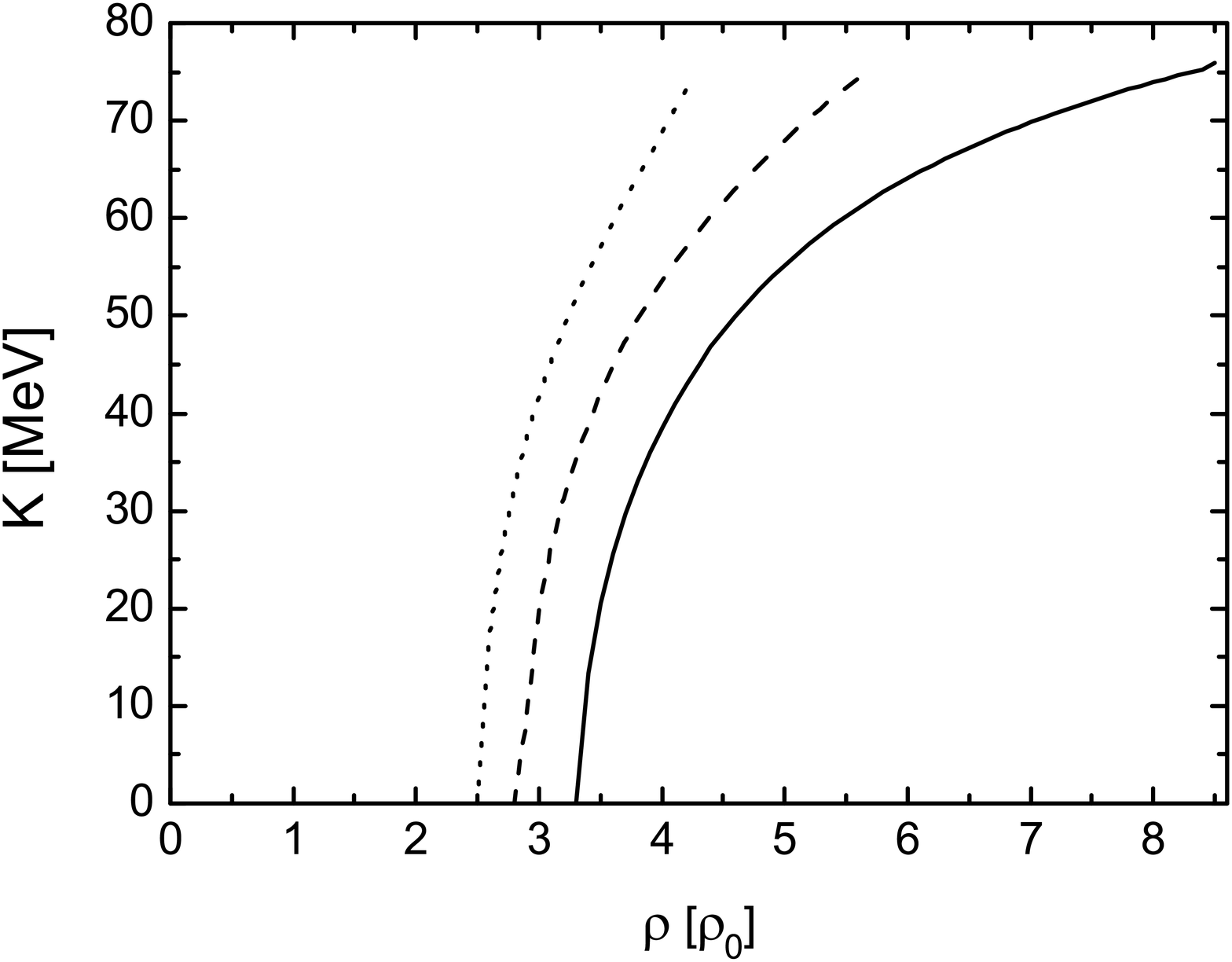}
\caption{The kaon amplitude for the LCK model with $\eta=-1$. See Fig.~\ref{chemsbal1} for the different lines.}
\label{kaonamps1}
\end{center}
\end{figure}

\begin{figure}[t!]
\begin{center}
\includegraphics[width=8.6cm]{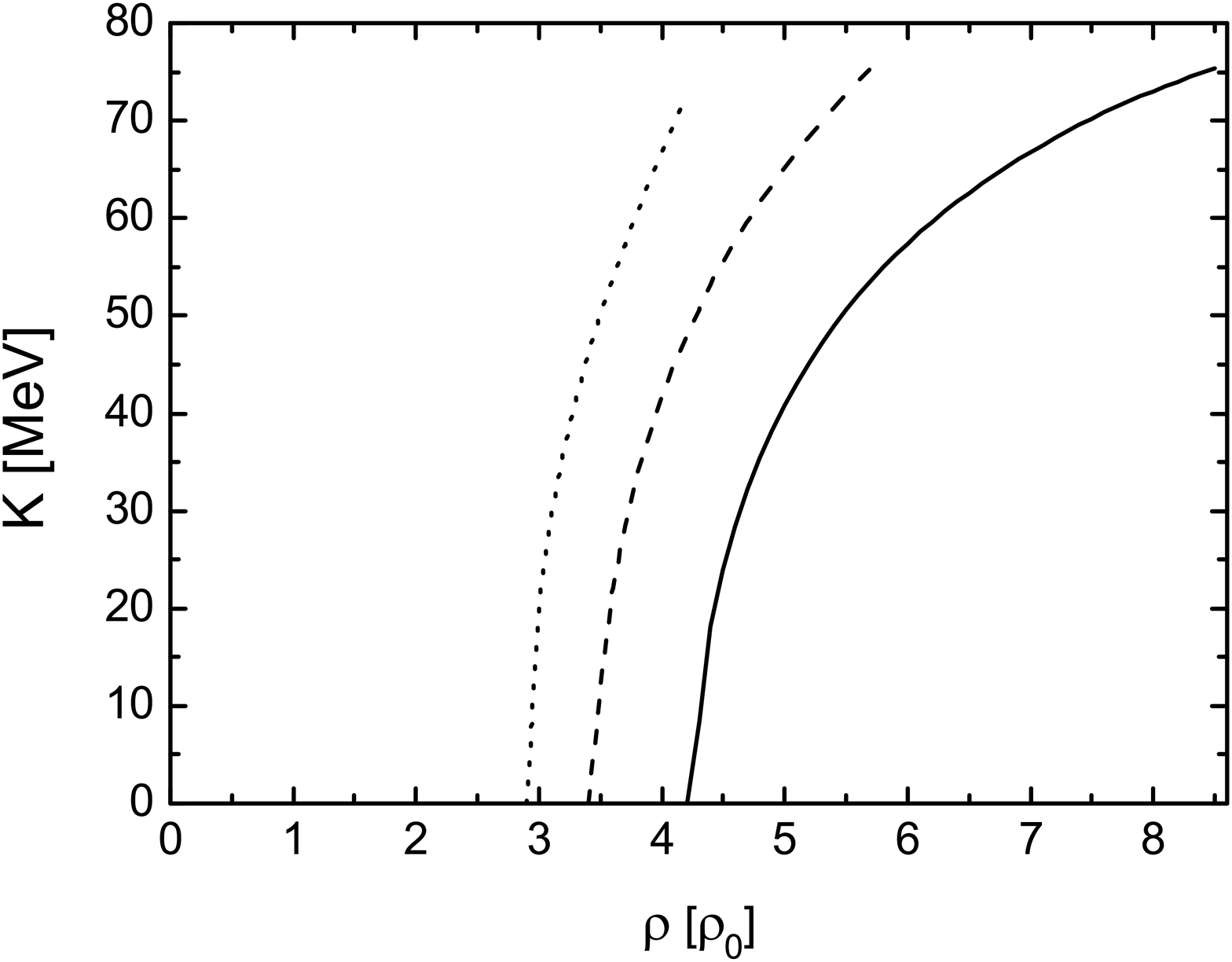}
\caption{The kaon amplitude for the LCK model with $\eta=0$. See Fig.~\ref{chemsbal2} for the different lines.}
\label{kaonamps2}
\end{center}
\end{figure}

As stated above, there is only one parameter left undetermined in the kaonic sector. We vary it in what we consider to be a reasonable range, say, $200 {\rm MeV}\lsim \Sigma_{KN}< 400$ MeV and see how the NM EoS is changed with kaon condensation. The density dependence of $\mu$ is shown in Figs.~\ref{chemsbal1} and \ref{chemsbal2} for $\Sigma_{KN}=200,300, 400$ MeV. In the NM-KNM system, the chemical potentials follow these lines. The dash-dotted line denotes that of the NM system.
It can be seen that the chemical potential $\mu$ begins to drop at the kaon threshold density, $\rho_t$, where the kaon amplitude $K$ starts developing non-zero value as shown in Figs.~\ref{kaonamps1} and \ref{kaonamps2}. We can also see that $\mu$ or equivalently the in-medium effective kaon mass $m_K^*$ vanishes at some high density, which we refer to as critical density denoted $\rho_c$, which could be identified with the chiral restoration density.

The contributions to the energy density and the pressure from kaon condensation are given by
\be
\epsilon_K &=& \left( m_K^2 + \mu_K^2 - \frac{\rho}{f^2} \Sigma_{KN} \right) K^2, \label{ek} \\
P_K &=& - \left(m_K^2 - \mu^2\right) K^2. \label{pk}
\ee
One can see that the kaon condensation gives a negative contribution to the total pressure for $\mu_K < m_K$. The energy density and the pressure of the system are given by
\be
\epsilon &=& \tilde{V}(\rho) + \rho (1- 2 \frac{\rho_p}{\rho})^2 S(\rho) + \epsilon_{lepton} \nonumber\\
&+& \Theta(K)\epsilon_K, \label{nmknmene}\\
P &=& \rho^2 \frac{\partial V(\rho)/\rho}{\partial \rho} +
\rho^2(1-2\frac{\rho_p}{\rho})^2\frac{\partial S(\rho)}{\partial \rho} + P_{lepton} \nonumber\\
&+& \Theta(K) P_K, \label{nmknmpre}
\ee
where $x=\rho_p / \rho$.
\begin{figure}[t!]
\begin{center}
\includegraphics[width=8.6cm]{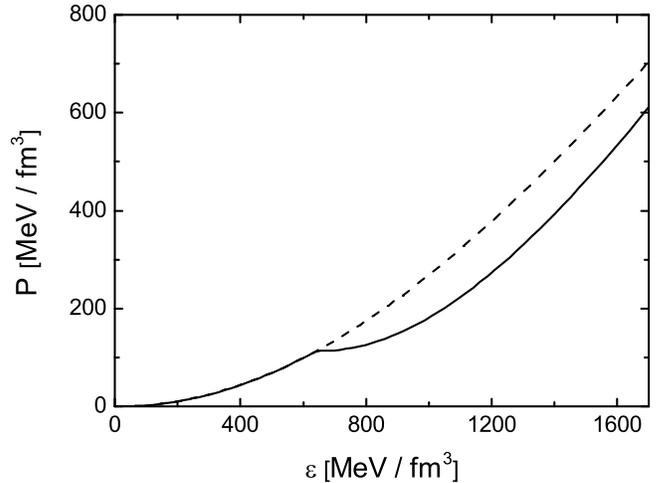}
\caption{The EoS for the NM (dashed line) and KNM (solid line) with $\eta=-1$ and $\Sigma_{KN}=130\MeV$.}
\label{eosnmknmbal1}
\end{center}
\end{figure}

\begin{figure}[t!]
\begin{center}
\includegraphics[width=8.6cm]{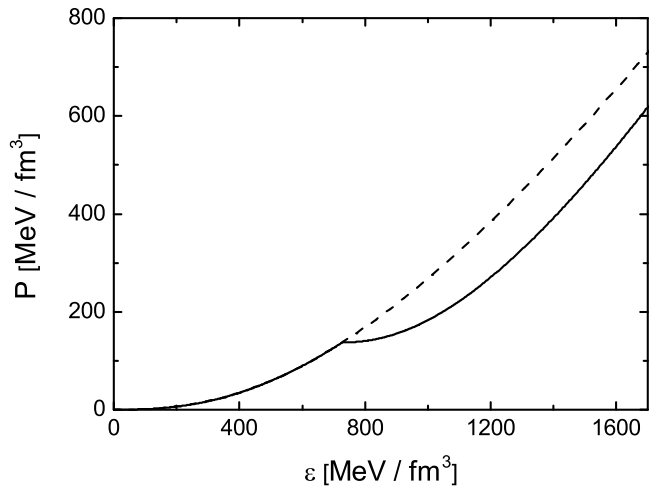}
\caption{The EoS for the NM (dashed line) and KNM (solid line) with $\eta=0$ and $\Sigma_{KN}=200\MeV$.}
\label{eosnmknmbal2}
\end{center}
\end{figure}

To have a qualitative idea of what's going on, we pick $\Sigma_{KN}\sim 200$ MeV for $\eta=0$ and $\Sigma_{KN}\sim 130$ MeV for $\eta=-1$. We do this because it is the minimum value at which the slope of the pressure vs. energy density remains positive, so that we can avoid resorting to Maxwell or Gibbs construction.  We see that the results so obtained are not physically reasonable; a higher $\Sigma_{KN}$ seems to be required to be compatible with nature. {We list the kaon threshold density $\rho_t$, critical density $\rho_c$, and chemical potential at the kaon threshold $\mu(\rho_t)$ for given $\Sigma_{KN}$'s in Table 2}.
\begin{table}
\begin{center}
\begin{minipage}{.4\textwidth}
\begin{center}
\begin{tabular}{c | c | c | c | c}
\hline
\hline
Model & $\Sigma_{KN}$ & $\rho_t$ & $\rho_c$ & $\mu(\rho_t)$\\
\hline
$\eta=-1$   & 130 & $3.74$   & $13.20$   & $281.9$\\
\hline
$\eta=0$   & 200 & $4.27$   & $8.59$   & $220.8$\\
\hline
\hline
\end{tabular}
\caption{The kaon threshold and critical densities for $\eta=-1$ and $\eta=0$ and the $\Sigma_{KN}$'s used for the TOV calculations (for details see next section). $\rho_t$ and $\rho_c$ are given in $\rho_0$ unit, and $\Sigma_{KN}$ and $\mu(\rho_t)$ are given in $\MeV$.}
\end{center}
\end{minipage}
\end{center}
\label{dens}
\end{table}

In Figs.~\ref{eosnmknmbal1} and \ref{eosnmknmbal2} are plotted the EoS for $\Sigma_{KN}\sim 130\MeV$ and $\sim 200\MeV$.
We see that the EoS is softened, as expected, beyond the kaon threshold density, $\rho_t$.

\subsection{Strange Quark Matter (SQM) Driven by Kaon Condensation}
We now turn to the possibility that kaon condensation can drive the dense system to a SQM at the critical density $\rho_c$ defined by the condition, $\mu_K =0$. For this, we assume, for simplicity, massless $u$, $d$, and $s$ quarks. Eq.~(\ref{musqm}) implies that
they have the same number densities\footnote{In a more realistic calculation we have
to take into account of strange quark mass and electrons and/or muons: $\rho_u =\rho_d \neq \rho_s$.},
\be
\rho_u =\rho_d =\rho_s =\rho_Q
\ee
Then the charge neutrality
\be
\frac{2}{3}\rho_u - \frac{1}{3}\rho_d - \frac{1}{3}\rho_s = 0
\ee
is automatically satisfied and there is no  need for additional leptons.
Kaon condensed nuclear matter will naturally go over to the SQM in $SU(3)$ symmetric phase in the massless limit.
The EoS of SQM is then given by
\be
\epsilon_{SQM} &=& a_4\frac{9}{4\pi^2}\mu_q^4 + B ~~ = ~~ 4.83a_4\rho^{4/3} + B,\\
P_{SQM}        &=& a_4\frac{3}{4\pi^2}\mu_q^4 - B ~~ = ~~ 1.61a_4\rho^{4/3} - B,
\ee
where $B$ is the bag constant~\cite{mitbag}. {Here $a_4$ denotes the perturbative  QCD correction~\cite{LP-GFEST,lapr}, which takes the value  $a_4 \leq 1$. The equality holds for SQM without QCD corrections.}
In the example of the previous section with $\Sigma_{KN}$= 200 MeV (for $\eta=0$), without perturbative QCD corrections (i.e., $a_4=1$), the $\rho_c$ comes out to be too high so that the bag constant that satisfies the phase boundary matching condition
\be
P_{KNM}(\rho_c) =  P_{SQM}(\rho_c^Q) \label{pmatch}
\ee
becomes negative.  Thus, for smaller values of $\Sigma_{KN}$, the kaon-driven scenario may not work.

If we instead take $\Sigma_{KN}\simeq 259$ MeV, the pressure matching can be satisfied. And we get the critical density $\rho_c= 6.37\rho_0$ both for $\eta=0$ and $-1$. Using chemical equilibrium conditions we get the critical quark number densities, $\rho^Q_c= 19.26\rho_0$ and  $19.11\rho_0$ respectively. One can see that $\rho^Q_c/3$ are not much different from $\rho_c$, as shown in Table \ref{4}. Given the critical densities, we can find a possible set of parameters, $a_4$ and $B^{1/4}$. For example, with $a_4 = 0.62$, we get $B^{1/4}=94.71\MeV$ and $97.25\MeV$ for $\eta=0$ and $\eta=-1$, respectively.

\begin{figure}[t!]
\begin{center}
\includegraphics[width=8.6cm]{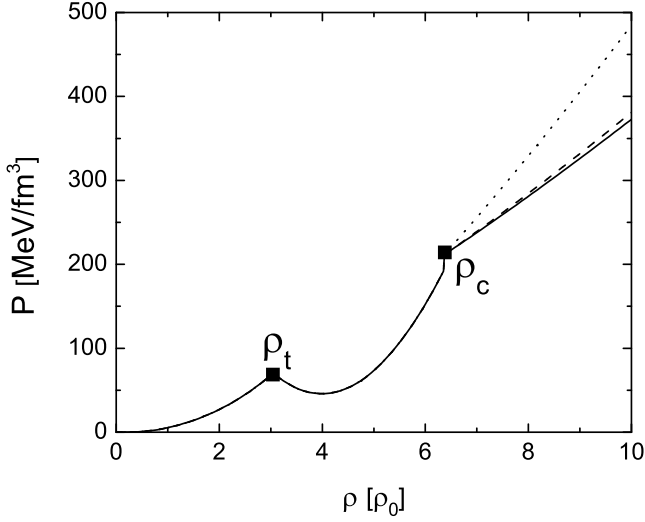}
\caption{The phase diagram of NM-KNM-SQM for LCK with $\eta=-1$. The solid, dased, and dotted lines are corresponding to $a_4$ = 1.0, 0.62, and 0.59, respectively.}
\label{nkqpre_bal1}
\end{center}
\end{figure}

\begin{figure}[t!]
\begin{center}
\includegraphics[width=8.6cm]{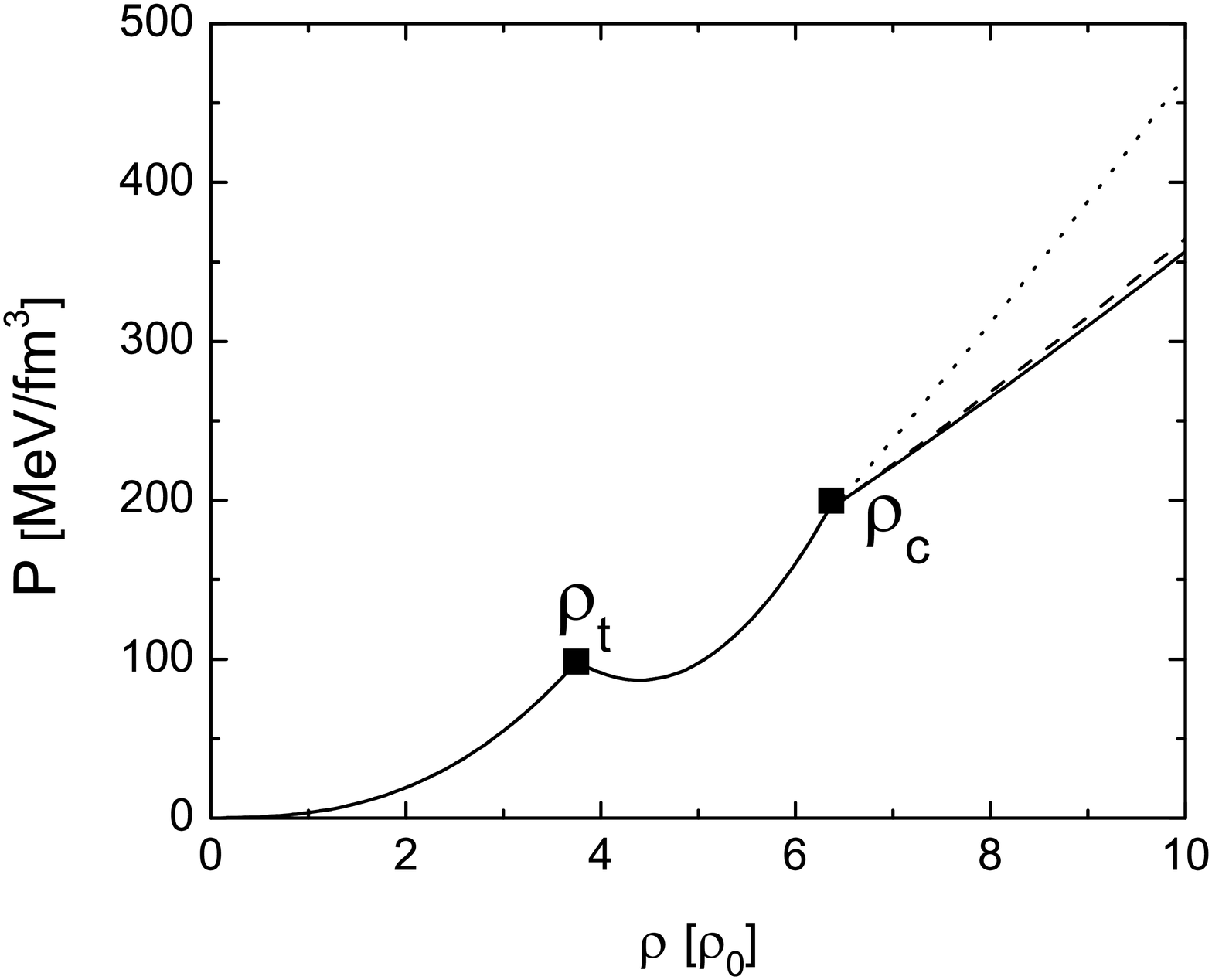}
\caption{The phase diagram of NM-KNM-SQM for LCK with $\eta=0$. The solid, dased, and dotted lines are corresponding to $a_4$ = 1.0, 0.62, and 0.59, respectively.}
\label{nkqpre_bal2}
\end{center}
\end{figure}

The different choice of bag constant with different parameters for symmetry-energy, $\eta$, and QCD corrections, $a_4$, is attributable to the different EoS depending on $\eta$ and $a_4$ which results in the different  pressures $P_{KNM}(\rho_c)$ for the boundary matching condition, Eq.~(\ref{pmatch}).
In Figs.~\ref{nkqpre_bal1} and \ref{nkqpre_bal2}, we plot the pressure profiles and tabulate the densities, $\rho_t$, $\rho_{t'}$, and $\rho_c$ in Table \ref{4}.  Here  $\rho'_t (\neq \rho_t)$ is  defined by
\be
P_{NM}(\rho_t) = P_{KNM}(\rho_{t'}).
\ee


\begin{table}[h]
\begin{center}
\begin{minipage}{.4\textwidth}
\begin{center}
\begin{tabular}{c | c | c | c | c}
\hline
\hline
Model & $\rho_t$  &   $\rho_{t'}$  & $\rho_c$ & $\rho^Q_c / 3$\\
\hline
$\eta=-1$ & $3.05$ & $4.95$ & $6.37$ & $6.37$\\
\hline
$\eta=0$ & $3.76$ & $5.03$ & $6.37$ & $6.42$\\
\hline
\hline
\end{tabular}
\caption{The characteristic densities of NM-KNM-SQM with the LCK models for $\eta=-1$ and $\eta=0$ with $\Sigma_{KN} \simeq 259 \MeV$. The values in the last column are the quark critical densities which are equivalent to the baryon number densities, respectively.}\label{4}
\end{center}
\end{minipage}
\end{center}
\end{table}

\section{Stellar Structure With Kaon Condensation}
We now turn to a stellar object that has an EoS with strangeness. Here we discuss the case where kaon is condensed but without SQM and the case where kaon condensed and  with strange quark core.
\subsection{Double-layered (NM+KNM) structure}
\begin{figure}[t!]
\begin{center}
\includegraphics[width=8.6cm]{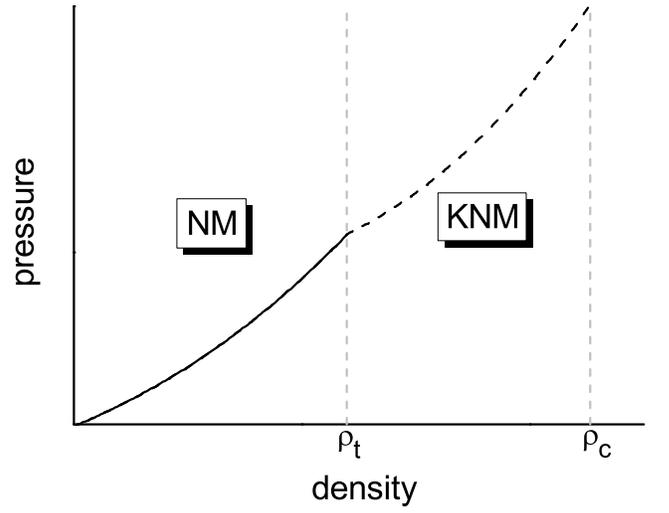}
\caption{The schematic phase diagram for NM-KNM.}
\label{nkpre}
\end{center}
\end{figure}

First we consider a simpler case, where neither Maxwell nor Gibbs construction is needed at the kaon condensation phase transition as shown in  a schematic $P$ vs $\rho$ curve (Fig.~\ref{nkpre}).  In the framework of our model with the LCK parametrization for EoS, the maximum values for $\Sigma_{KN}$ turn out to be $\sim 130$ MeV and $\sim 200$ MeV for $\eta=-1$ and 0 respectively.   For those values of $\Sigma_{KN}$, we cannot have pressure balance for SQM unless we allow a negative bag constant.  Hence up to the critical densities in these parameters, one finds NM and KNM from the outer layer to the core part.
We use Eqs.~(\ref{nmknmene}) and (\ref{nmknmpre}) with Eqs.~(\ref{ek}) and (\ref{pk}) to integrate the TOV equation
\be
\frac{dM}{dr} &=& 4 \pi \epsilon r^2, \nonumber \\
\frac{dP}{dr} &=& - \frac{G M \epsilon}{r^2} \left(1 + \frac{P}{\epsilon}\right) \left(1 + \frac{4 \pi r^3 P}{M}\right)
\left(1 - \frac{2GM}{r}\right)^{-1}, \label{toveqn}
\ee
where $M(r)$ is the mass  enclosed inside the  radius $r$.
\begin{figure}[t!]
\begin{center}
\includegraphics[width=8.6cm]{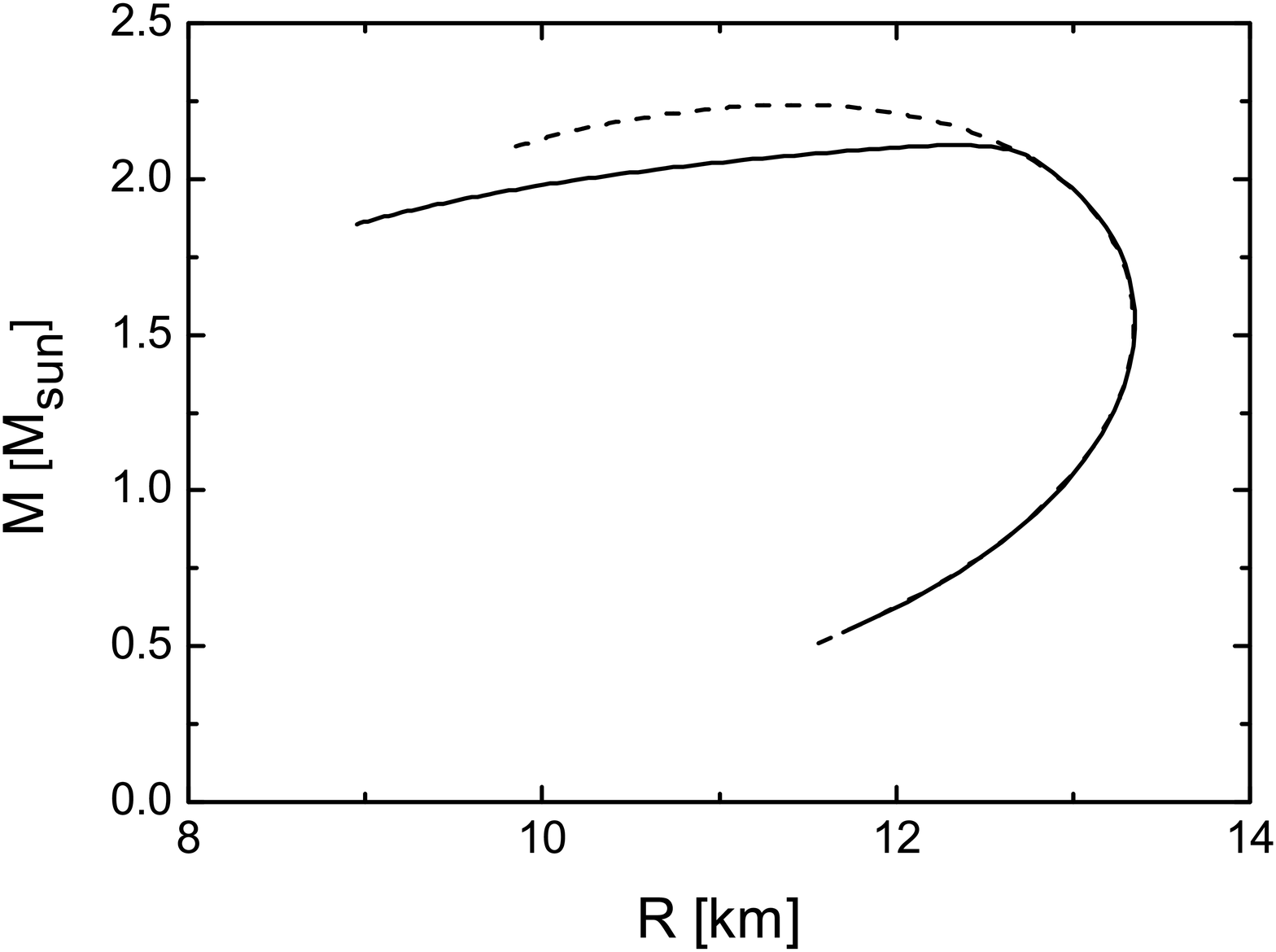}
\caption{ $M$-$R$ sequence for LCK with $\eta=-1$ and $\Sigma_{KN} = 130\ \MeV$. The solid and dashed lines are for the cases with  NM-KNM and NM, respectively.}
\label{rmnmknmbal1}
\end{center}
\end{figure}
\begin{figure}[t!]
\begin{center}
\includegraphics[width=8.6cm]{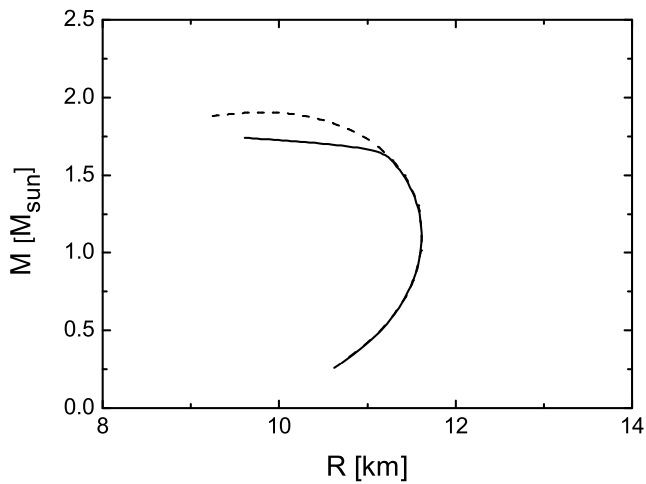}
\caption{ $M$-$R$ sequence for LCK with $\eta=0$ and $\Sigma_{KN} = 200\ \MeV$. The solid and dashed lines are for the cases with  NM-KNM and NM, respectively.}
\label{rmnmknmbal2}
\end{center}
\end{figure}

The results are given in Figs.~\ref{rmnmknmbal1} and \ref{rmnmknmbal2}.
They illustrate how softening the EoS with kaon condensation affects the mass and the size: It lowers the mass and increases the size from the case of NM at the same central density. However, because we use the relatively small values for $\Sigma_{KN}$, the effect of kaon condensation is not so significant (e.g., about $\sim 0.2M_\odot$ reduction in the maximum mass).

In Table~\ref{knmstellar} are summarized the stellar structure parameters of our results for both the NM and the NM-KNM systems. It is interesting to note that the two parametrizations for the symmetry energy, with $\eta=0$ and -1, give rather different features of $M$ vs. central density relations.   For the softer symmetry energy, $\eta=0$,  the mass $M$ increases with the central density beyond the kaon condensation threshold density going up to the critical density. It implies that the star is  gravitationally stable up to the critical density. Beyond the critical density,  we have no models to calculate the EoS. So we shall simply take the maximum mass in this case to be given by the mass with the critical density, $\rho_c$,  of Table \ref{4} taken as the central density of the star. However, with the stiffer symmetry energy with $\eta=-1$,  the maximum mass is obtained by the gravitational instability condition at the central density much smaller than the critical density but not far from the kaon threshold density. This feature can be seen in Table~\ref{knmstellar}.
\begin{table}[h]
\begin{center}
\begin{minipage}{.4\textwidth}
\begin{center}
\begin{tabular}{l | c | c | c}
\hline
\hline
Model & $M_{max}$ & $R|_{{}_{M=M_{max}}}$ & $\rho^{(c)}|_{{}_{M=M_{max}}}$\\
\hline
NM ($\eta=-1$)  & $2.23$ & $11.41$ & $6.1$ \\
\hline
NM+KNM ($\eta=-1$)   & $2.10$ & $12.33$   & $5.2$\\
\hline
NM ($\eta=0$)  & $1.90$ & $9.81$ & $8.1$ \\
\hline
NM+KNM($\eta=0$)   & $1.74$ & $9.61$   & $8.5$\\
\hline
\hline
\end{tabular}
\caption{The calculated structure parameters for various systems (see text for details). The units are $M_{\odot}$, km, and $\rho_0$ for mass, size, and central density, respectively.}\label{knmstellar}
\end{center}
\end{minipage}
\end{center}
\end{table}

\subsection{Triple-layered (NM+KNM+SQM) structure}
\begin{figure}[t!]
\begin{center}
\includegraphics[width=8.6cm]{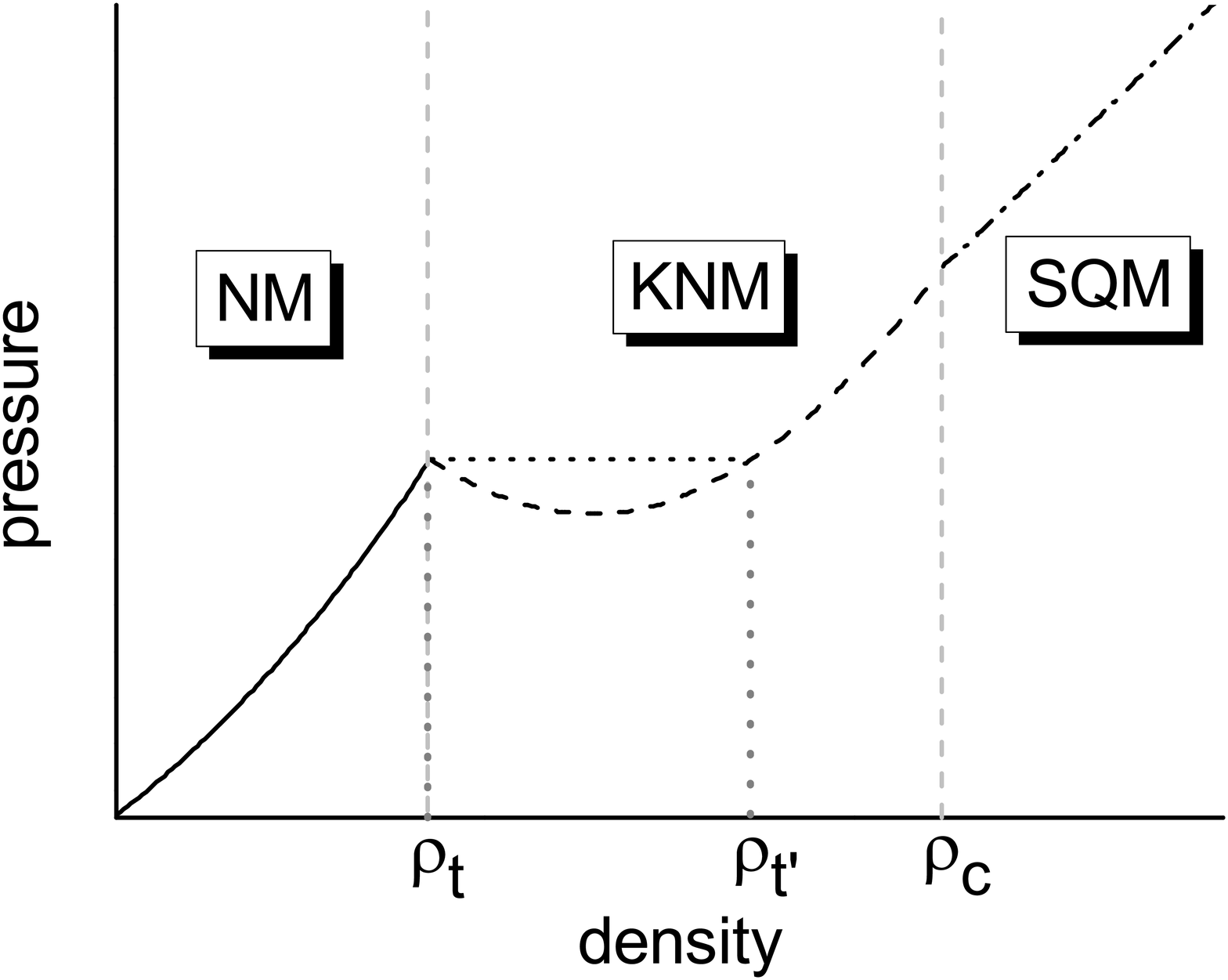}
\caption{The schematic phase diagram for NM-KNM-SQM.}
\label{nkqpre}
\end{center}
\end{figure}

\begin{figure}[t!]
\begin{center}
{\includegraphics[width=8.6cm]{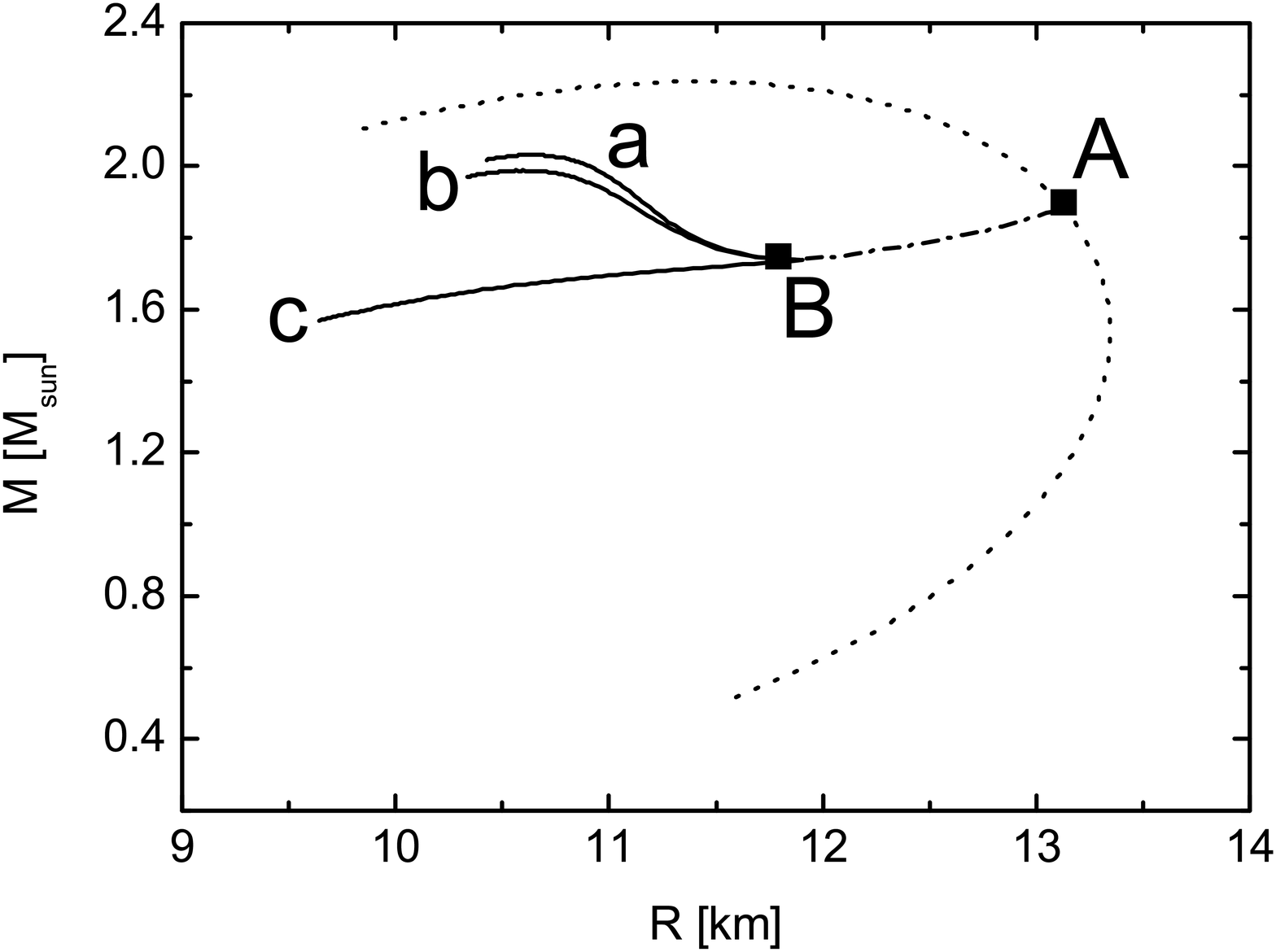}} \hfill
\caption{The M-R sequences for LCK with $\eta=-1$.  The dotted line denotes NM.  The dashed-dotted line between A and B denotes the double-layered (NM-KNM) system.  The solid lines, a, b and c,  denote the triple-layered (NM-KNM-SQM) system with the QCD corrections with $a_4 = 0.59, 0.62$ and $1$, respectively. }
\label{MR-}
\end{center}
\end{figure}

\begin{figure}[t!]
\begin{center}
{\includegraphics[width=8.6cm]{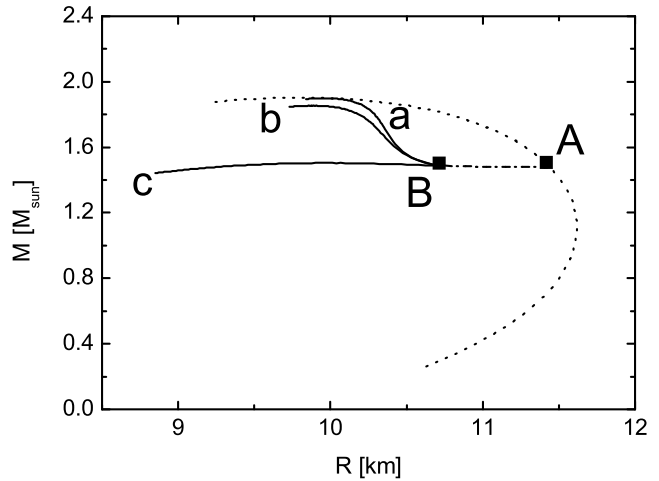}} \hfill
\caption{The M-R sequences for LCK with $\eta=0$.  The dotted and dashed-dotted lines are the same as in Fig.~\ref{MR-}. The solid lines, a, b and c,  denote the triple-layered (NM-KNM-SQM) system with the QCD corrections with  $a_4 = 0.59, 0.62$ and $1$, respectively. }
\label{MR0}
\end{center}
\end{figure}

If we take a larger value for $\Sigma_{KN}$, say, 260 MeV, it is possible to consider a triple-layered structure shown schematically in Fig.~\ref{nkqpre} consisting of NM, KNM, and SQM from the outer layer to the core part.
In contrast to the case considered above, there will be an unstable region between $\rho_t$ and $\rho_{t'}$, for which we have to resort to Maxwell or Gibbs construction. In the Maxwell construction, it turns out to be impossible to match the chemical potential, which signals instability.  In the Gibbs construction, however, there appears a mixed phase of NM and KNM. In this work,  we take a rather simple approach, namely, allow the discontinuity of density and chemical potential by assuming that NM with the density $\rho_t$  changes into KNM with  the density $\rho'_t$ at the phase boundary defined by
\be
P(\rho_t) = P(\rho_{t'}).
\ee
The resulting mass-radius relations that follow from the TOV equation are plotted in Figs.~\ref{MR-} and \ref{MR0}. As expected, kaon condensation with higher $\Sigma_{KN}$'s would lead to smaller masses. However, when the central density becomes higher, an SQM driven by kaons appears at the core. No sharp change due the emergence of SQM is observed, which implies that the kaon-driven SQM transition is a rather smooth transition in this scenario.

\begin{table}[h]
\begin{center}
\begin{minipage}{.4\textwidth}
\begin{center}
\begin{tabular}{c | c | c | c | c}
\hline
\hline
Model & $a_4$ & $M_{max}$ & $R|_{{}_{M=M_{max}}}$ & $\rho^{(c)}|_{{}_{M=M_{max}}}$\\
\hline
$\eta=-1$   & 0.62 & $1.99$ & $10.60$ & $12.3$\\
\hline
$\eta=0$   & 0.62 & $1.85$ & $9.86$ & $13.5$\\
\hline
\hline
\end{tabular}
\caption{The maximum masses, $M_{max}$, radii, $R$, and central densities, $\rho_{(c)}$ for the three-layered structure  with the QCD corrections, $a_4 = 0.62$ for both of $\eta = -1$ and $0$ respectively. The units of the stellar structure parameters are the same as in Table \ref{knmstellar}.}\label{dens1}
\end{center}
\end{minipage}
\end{center}
\end{table}

The results given in Table \ref{dens1} highlight the importance of perturbative QCD corrections in viable three-layered structure.
We find that with the EoS of SQM without perturbative  corrections, i.e., $a_4=1$, the results (the solid line c in Figs.~\ref{MR-} and \ref{MR0}) are  similar to the previous case with the soft symmetry energy (with $\eta=0$) of the kaon condensed matter, which is present up to the critical density. Here the mass increases with the central density even beyond the kaon threshold density to reach the maximum  mass, which is slightly greater than the mass at the kaon threshold. However its detailed structure is quite different from the latter with $\Sigma_{KN}= 200$ MeV, where there is no room for the SQM. If, however, we include the QCD corrections, the maximum mass lies beyond the critical density, and the core is in the form of SQM.
With the stiffer symmetry energy with $\eta=-1$,  the maximum mass of $\sim 1.99 \msun$ can be obtained by the gravitational instability condition  at the central density $\rho= 12.3\ \rho_0$. With the softer symmetry energy (with $\eta=0$),  the maximum mass of $\sim 1.85 \msun$ can be  obtained  at the central density $\rho= 13.5\ \rho_0$, as shown in solid lines $b$ in Figs. 13 and 14 and in Table~\ref{dens1} with $a_4=0.62$. For different set of parameters with $a_4=0.59$, we get slightly larger maximum masses, $2.03 \msun$ and $1.90 \msun$, respectively, for $\eta=-1$ and $\eta=0$, as shown with solid lines $a$ in Figs. 13 and 14. This is a rather high density and a more sophisticated treatment of SQM than the simple MIT bag model with perturbative corrections (e.g., color superconductivity) may be called for and change the maximum mass.

\section{Summary and Discussion}

We have proposed a scenario in which dense compact-star matter is driven smoothly to an SQM  by kaon condensation at the density at which the kaon chemical potential $\mu_K=m_K^*$ becomes negligibly small.  It is  based on the main assumption of the paper that kaon condensation does occur at a finite density and that the kaon effective mass should become very small at a higher density, which, however, should not be larger than that of compact star core.  We show that the compact star with an NM-KNM-SQM structure  is possible with the parameters that are not excluded by theory or phenomenology.  We admit that our calculation is highly model-dependent without the benefit of constraints by QCD or by reliable models, but it is simple and could be easily falsified by experiments. All three states of matter, that is, NM, kaon condensed matter, and QM, are treated on the same level of accuracy. For kaon-nuclear interactions, we take the simplest chiral effective Lagrangian at tree order, which seems to be supported by the available in-medium lattice QCD with the $\Sigma_{KN}$ term being taken as the only unknown quantity. As for nuclear interactions, we take LCK's empirically parameterized form for the energy density and symmetry energy. There are potentially serious effects that are missing in the treatments and that could drastically change the scenario. They are discussed in Appendix.

\begin{figure}[t!]
\begin{center}
\includegraphics[width=8.6cm]{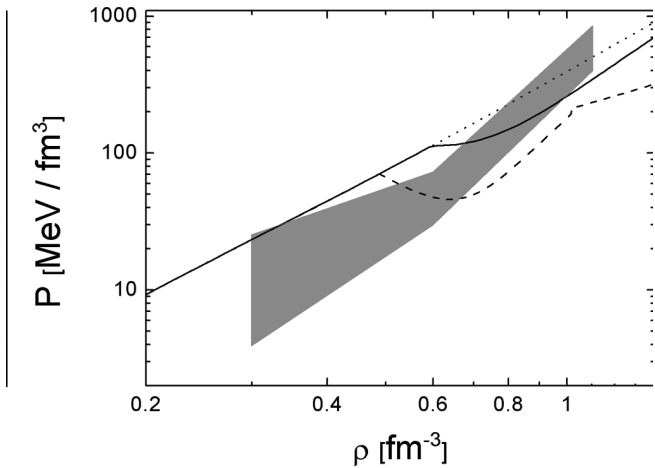}
\caption{Pressure vs. density ($\rho$) for NM-KNM (solid line) and NM-KNM-SQM (long-dashed line) with LCK ($\eta=-1$) and considered parameters in previous sections compared with the analysis by \"{O}zel {\it et al.} (shaded area)\cite{ozel}. The dotted line denotes NM, which is given for comparison with NM-KNM and NM-KNM-SQM.}
\label{logprebal1}
\end{center}
\end{figure}
\begin{figure}[t!]
\begin{center}
\includegraphics[width=8.6cm]{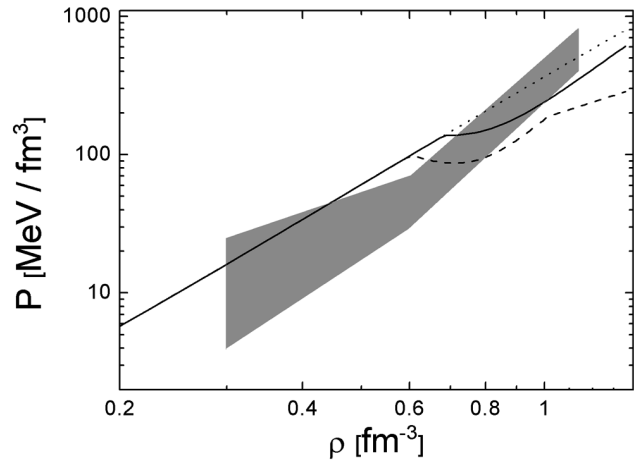}
\caption{The same as Fig.~\ref{logprebal1} for LCK ($\eta=0$).}
\label{logprebal2}
\end{center}
\end{figure}

We find that the maximum mass of compact stars without kaon condensation with the LCK parameters can be twice the solar mass, which is consistent with the recent observation.  With kaon condensation, the EOS becomes soft and the maximum mass gets reduced, as expected.   With $\Sigma_{KN}= 200$ MeV (and $\eta=0$), an SQM driven by kaon condensation in our scenario is impossible because the pressure matching at the SQM boundary gives a negative bag constant.   However with the somewhat larger $\Sigma_{KN} = 260$ MeV,  an SQM driven by kaon condensation can be formed with the bag constants $B^{1/4} \simeq 188$ and 179 MeV, respectively, for $\eta=-1$ and $\eta=0$.\footnote{Note that for the case of the kaon-condensation-driven SQM, $\Sigma_{KN} \simeq 259$ MeV is fixed for $\eta$'s.  However, the bag constants need to be different to match the pressure condition.} {The maximum mass can be as large as $2 \msun$ for $\eta=-1$ at A in Fig.~\ref{MR-}, where only NM is relevant. For $\eta=0$, a triple-layered structure is possible, but without perturbative QCD corrections in SQM, the maximum mass is far below $2 \msun$, as given by the solid line c in Fig.~\ref{MR0}.} When suitable QCD corrections are included in the EoS of SQM, smaller bag constants are required,\footnote{The bag constants needed for the case with perturbative corrections may appear to be too low in comparison with the values needed in vacuum phenomenology. This may not be a cause for worry, however. Because the part of the gluon condensate locked to the quark condensate is expected to ``melt" across the chiral transition point, the smaller bag constant effective in the SQM phase appears to be more natural.} $97.25$ and $94.71$ MeV, respectively, for $\eta=-1\ (a_4=0.62)$ and $\eta=0\ (a_4=0.62)$. It is also found that the transition is smooth as far as the mass change is concerned. This indicates that the kaon condensation as we described here can be considered a ``doorway" to a quark-star matter in the core.

What we learn from our analysis made in this paper is that the nature (stiffness or softness) of the EoS in the nucleon sector represented in the LCK model by the different $\eta$ values (even within the range allowed by experiments), the sigma term $\Sigma_{KN}$, the bag constant, and QCD corrections is highly correlated. Indeed, for the stiffer EoS with $\eta=-1$ and the smaller $\Sigma_{KN}\sim 130$ MeV, a two-layer structure of NM and  KNM can be made compatible with the mass vs radius constraint given by the 1.97 $M_{\odot}$ star.   With the larger $\Sigma_{KN} \simeq 260$ MeV,  even the three-layer structure of NM, KNM, and SQM with an appropriate bag constant is possible beyond the critical density, $6.37 \rho_0$. And perhaps most significantly, the maximum mass of a neutron star with the three-layer structure can be made consistent with the recent observation.  There is, of course, nothing new or surprising in this sort of correlations in nuclear processes. However, it points for a reliable treatment of the $M$ vs. $R$ ratio to the necessity of treating {\em all} of the variety of different aspects of the phenomenon on the same footing. This point can be illustrated by the following observation. With a relatively stiff EoS with nucleons only such as that typified by the one constructed in Ref. \cite{APR} (which is comparable to the LCK's with $\eta=-1$), the critical density for kaon condensations can be pushed to too high a value to be of relevance to the process, say, $\gsim 7\rho_0$~\cite{pandha}. However, as discussed in the Appendix, an effective field theory consideration based on hidden local symmetry suggests that at a certain high density approaching the density regime where quark degrees of freedom emerge, the repulsion that pushes up the critical density for kaon condensation may get strongly suppressed as one approaches the critical density. This means that the NM and KNM sectors have to be treated in the same framework. Similar remarks can be made for the SQM sector.

In our treatment, strangeness was brought in by kaons. It could have been done by hyperons. As argued in Section \ref{strategy}, the two ways at the mean-field level should capture approximately the same physics. They may not be representing alternatives. We expect that they will give qualitatively same effects on the stellar structure.

The results we obtained are compared with the analysis by \"{O}zel, et al. \cite{ozel} shown  in Figs.~\ref{logprebal1} and \ref{logprebal2} as shaded areas. In view of the variety of intricate mechanisms left out in our treatment, it is not obvious that one can do much better with more detailed model calculations. As stressed, what is needed is a consistent field theoretic approach that can account also for the effects of the type discussed in Appendix. We are only at the beginning of such an endeavor.

\section*{Acknowledgments}
We are grateful for useful and enlightening comments from Jim Lattimer on the matter discussed in this paper. This work is supported by WCU (World Class University) program: Hadronic Matter under Extreme Conditions through the National Research Foundation of Korea funded by the Ministry of Education, Science and Technology (R33-2008-000-10087-0). KK is partially supported by the Creative Research Program (2011) of the National Institute for Mathematical Sciences (NIMS) in Korea.

\setcounter{equation}{0}
\renewcommand{\theequation}{A\arabic{equation}}
\section*{Appendix}
In this appendix, we address a variety of issues that are not taken into account in our work: The role of $\Lambda (1405)$ in dense kaonic nuclei and kaon condensation, topological effects on the kaon propagation in dense matter and possible binding of kaons in finite and infinite systems, the fate of $\omega$-meson-mediated repulsion at high density. These issues are potentially relevant to the problem addressed in our paper. To make a reliable treatment taking all these considerations into account would require a fully consistent field theoretic treatment and such a treatment is presently lacking in the literature. Most of the treatments found in the literature are a piece-meal patching of different ingredients with no obvious connection between them. We are unable to do much better here either. What we can do is just to point out how the various ingredients can figure in kaon-nuclear and high density physics. In so doing, however, we have in mind a specific field theoretic framework based on HLS~\cite{HY:PR} and its generalized form  -- which by itself is consistent but incomplete -- although we do not have the formulation to allow us to do the calculation explicitly within that framework.

\subsection*{(A) The role of $\Lambda (1405)$}
In considering the possible mechanisms for or against the formation of dense kaonic (finite) nuclei, the $\Lambda (1405)$ has figured prominently. There is a humongous controversy on whether and how $\Lambda (1405)$ enters in generating strong or weak $\Lambda (1405)$-nuclear potential. We gloss over this controversy here because it does not seem to be essential to our approach. What we would like to describe here is that, regardless of what the $\Lambda (1405)$ does near the threshold $K^-$-nucleon interactions and how it generates the  $\Lambda (1405)$-nuclear potential, it is irrelevant to our approach to kaon condensation. If one were to try to arrive at the putative kaon condensation by starting from elementary kaon-nucleon interactions, the structure of the $\Lambda (1405)$ would seem to matter. In fact, it has been suggested that even if there is a strongly attractive kaon-nuclear interaction, kaons do not condense macroscopically~\cite{gal}. There is a difference in perspectives on this issue.

\begin{figure}[t!]
\begin{center}
\includegraphics[width=8.6cm]{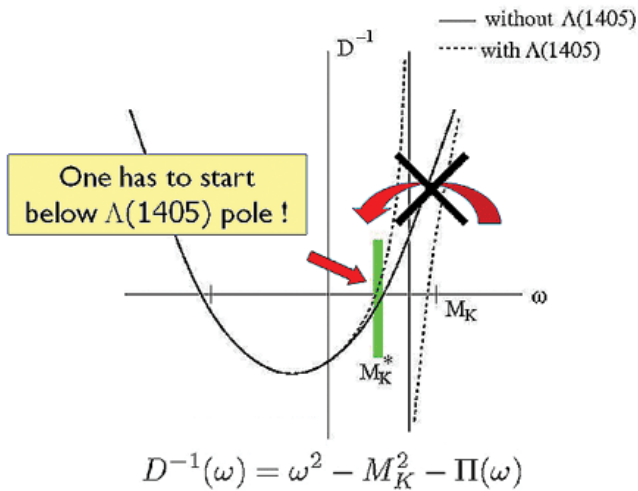}
\vskip -1.cm
\caption{(Color online) Irrelevance of $\Lambda (1405)$ in kaon condensation. $M_K^*$ is the effective kaon mass. }
\label{lambda}
\end{center}
\end{figure}

In Sec. \ref{kaoncon}, in addressing kaon condensation, we made no account of the presence of $\Lambda (1405)$. This is because, as was first pointed out in Ref. \cite{lmr}, the precise location of $\Lambda (1405)$ is irrelevant to the phase transition involved. This point is depicted schematically in Fig.~\ref{lambda}.\footnote{We are grateful to Chang-Hwan Lee for allowing us to use this figure.}

There $D^{-1} (\omega)$ gives the dispersion formula for the kaon frequency $\omega$.\footnote{We are, of course, assuming that we can treat $\Lambda(1405)$ as a quasiparticle in medium. If this were not the case, the spectral function could be more complicated. Even so, the qualitative feature of our argument will still hold as suggested by the RG argument of Ref. \cite{LRS}.} Given that kaon condensation occurs far below the threshold, that is, with $M_K^*$ far below the $\Lambda (1405)$ pole, the $\Lambda (1405)$-nuclear interaction is irrelevant in the RGE sense of Ref. \cite{LRS} so that it makes no difference whether $\Lambda (1405)$ is there or not. What this tells us is that kaon-nuclear interactions in dilute system have no influence on the presence of the phase transition, though they could influence its location.

\setcounter{equation}{0}
\renewcommand{\theequation}{B\arabic{equation}}
\subsection*{(B) Effect of topology change in dense matter on EoS}
\subsubsection*{(B1) Skyrmion-half-Skyrmion transition}
When dense matter is simulated by putting Skyrmions on crystal lattice, which is potentially a powerful tool for looking at strong-coupling dynamics of baryonic matter at high density~\cite{multifacet}, one discovers that there is a sort of phase transition at a certain density above NM density that we denote as $\rho_{1/2} > \rho_0$ from the Skyrmion matter -- which corresponds to the usual NM -- to a half-Skyrmion matter populated by half-soliton configurations. In that phase, chiral symmetry is putatively restored but quarks are still confined. Although the critical density $\rho_{1/2}$ cannot be pinned down, it presumably lies between 1.3 and 2 times the NM density. This transition is quite novel and has never been predicted before and there are no experimental data to confirm or refute this prediction. It could perhaps be vindicated or falsified in future experiments at such laboratories as FAIR/GSI, J-PARC, or RIB machines in construction.

There is a strong indication that hidden local symmetric Lagrangian, especially with an infinite tower as recently discovered, will be more reliable and more predictive than the Skyrme model with the pion field only for describing nuclear and dense matter~\cite{multifacet}.  No such treatment that is reliable is available to date. One reason why it can be taken seriously in the absence of a credible justification is that it involves a topology change. As such, its qualitative feature could be robust, more or less free of model dependence. In fact, this feature presents an advantage. Normal NM is most likely a liquid with its stability assured by the Fermi-liquid fixed point -- and not a solid. However, it is possible that the matter is solid at high density, as suggested by $N_c$ arguments. Now what we are observing on the crystal is a change of topology, so the properties that are anchored on topological structure could remain valid in what we are looking at although the density is not high enough.

That the Skyrmion-to-half-Skrymion transition takes place at some high density is more or less model-independent within the adopted framework, although detail features will depend on models. In Ref. \cite{PKR}, such a calculation was performed with the Skyrme Lagrangian implemented with the ``soft" dilaton that enters in the trace anomaly of QCD that is assumed to melt across chiral restoration~\cite{LR}. From this model, certain qualitatively robust features emerge. At $\rho_{1/2}$ where a Skyrmion fractionizes into two half-Skyrmions, the quark condensate $\la\bar{q}q\ra$ (in QCD variable) which is given in the Skyrme model by ${\rm Tr} U$ (where $U$ is the chiral field) vanishes {\em on the average} in one unit cell; thus, chiral symmetry apparently ``restored" but with a non-zero pion decay constant $f_\pi^*\neq 0$. In this phase, quarks are still confined in hadrons. Such a putative chiral symmetry restored but quark-confined phase has not been observed in continuum models. If confirmed to be viable, however, it will have an important impact on various phenomena, as described below.

\subsubsection*{(B2) Anomalous kaon binding}
\begin{figure}[ht]  
\centerline{
\begin{tabular}{c}
\epsfig{file=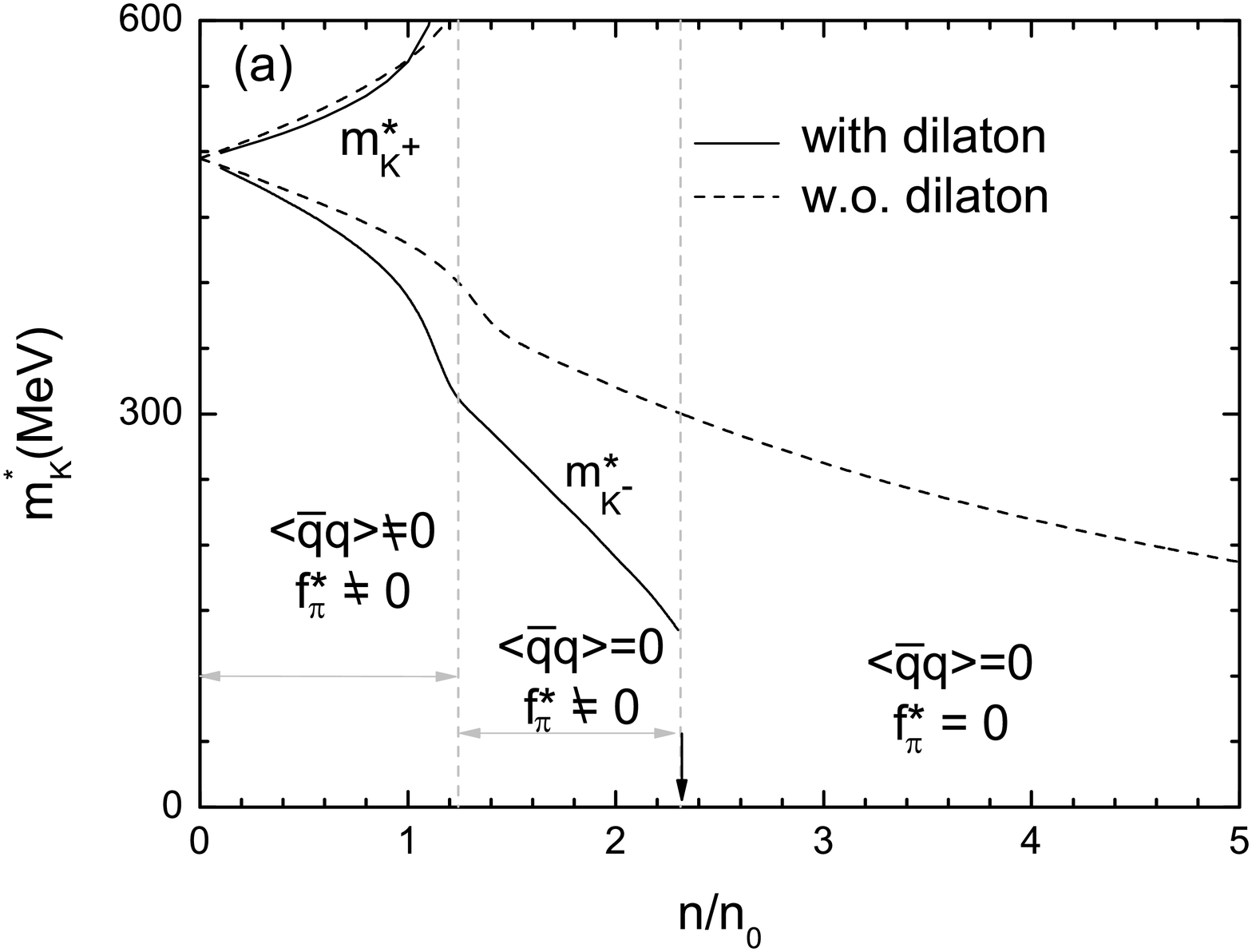, width=8cm, angle=0} \\
\epsfig{file=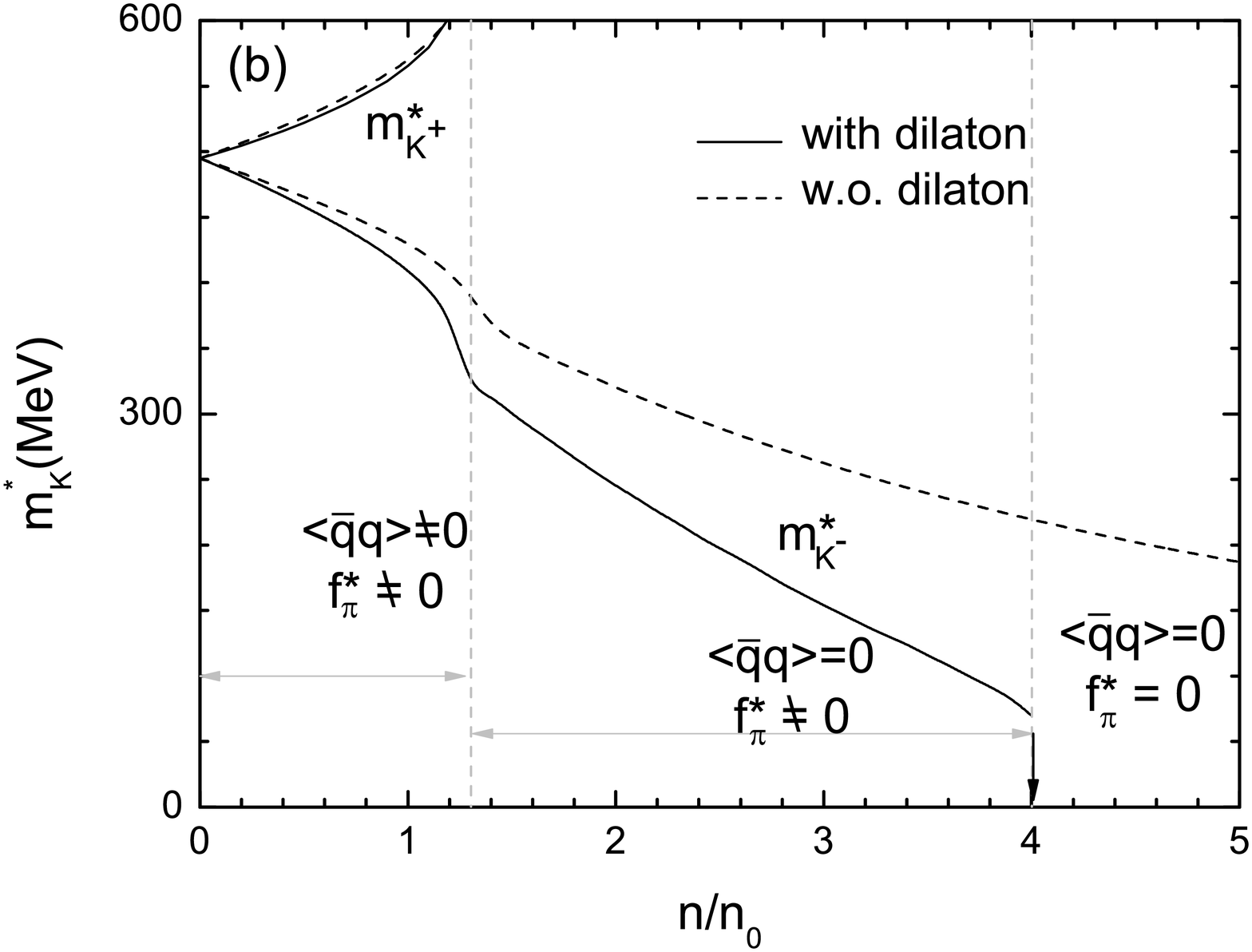, width=8cm, angle=0}
\end{tabular}
}
\caption{ $m^*_{K^\pm}$ vs. $\rho/\rho_0$  in dense skyrmion matter
which consists of three phases: (a) $\la\bar{q}q\ra\neq 0$ and
$f_\pi^*\neq 0$, (b) $\la\bar{q}q\ra=0$ and $f_\pi^*\neq 0$ and
(c) $\la\bar{q}q\ra=0$ and $f_\pi^*=0$. The parameters are fixed
at $\sqrt{2}ef_\pi=m_\rho=780$ MeV and dilaton mass $m_\chi=600$ MeV
(upper panel) and $m_\chi=720$ MeV (lower pannel). }
\label{fig1}
\end{figure}

One intriguing consequence of the topology change is ``anomalous" binding of $K^-$ in dense Skyrmion matter. As stressed above, a powerful treatment for dense baryonic matter could be made with a three-flavor effective Lagrangian with hidden local symmetry, but this is numerically involved and has not been worked out yet. Though less reliable, one can instead take the Skyrme Lagrangian as an effective one that results when all vector mesons are integrated out. Given such a Skyrmion Lagrangian, one can treat the property of a kaon in dense medium in terms of the kaon propagating in the background of Skyrmions put on a crystal lattice. This was worked out in \cite{PKR}. The main agent for binding the kaon to the Skyrmion in this description is the Wess-Zumino term in the $SU(3)_f$ Lagrangian. This attraction is the analog to the Weinberg-Tomozawa term in the chiral Lagrangian Eq.~(\ref{LKN}). However, there is an additional mechanism for kaon-nuclear attraction not apparent in the Lagrangian that comes from the Skyrmion-half-Skyrmion phase change. The result of Ref. \cite{PKR} is summarized in Fig.~\ref{fig1}, where the effective kaon mass $m_K^*$ is plotted as a function of density $\rho/\rho_0$ (denoted in the figure as $n/n_0$) for two different values of the (free-space) dilaton mass.\footnote{The precise value of the dilaton mass -- as well as the microscopic (QCD) structure of the dilaton -- is unknown. This is the part of the puzzle on the nature of scalar mesons in low-energy hadron physics. The values used here are what are considered to be reasonable values for the mass of the scalar that enters into mean field theory of NM.}  What we note is that the location of $\rho_{1/2}$ is remarkably insensitive to the dilaton mass, but $m_K^*$ varies strongly with density at $\rho_{1/2}$ and beyond. Thus the following warning: What is given in Fig.~\ref{fig1} should be taken with caution.

There are a several noteworthy observations to make on this result. One is that there is a sudden -- and sizable -- decrease in $m_K^*$ at $\rho_{1/2}$. Such an ``anomalous" drop cannot take place without topology change. Thus, it cannot be present in standard many-body or field theory treatments of nuclear and dense matter in which topology does not figure. What we learn from this observation is that if one were to produce a compact kaonic-nuclear system that can reach the density corresponding to $\rho_{1/2}$ -- which should be above, but not too high above, the normal matter density, there the system would undergo a drop in kaon mass that is not present in the standard treatment.\footnote{A similar anomalous drop is proposed in the work of Akaishi et al~\cite{yamazaki}. They referred to it as ``enhancement effect."} It is perhaps significant that a similar ``discontinuity" in nuclear tensor forces takes place at the same transition density $\rho_{1/2}$ mentioned below.

The other interesting observation is that the kaon mass drops down to a small value, with the kaon disappearing, at the point where $f_\pi^*$ falls to zero, presumably signaling the deconfined phase. In our scenario of kaon-condensation -- quark-matter continuity discussed in the text above, we are imagining an analogous process without, however, the apparent discontinuity of the magnitude that depends on the dilaton mass that is seen in Fig.~\ref{fig1}.

\subsection*{(B3) Tensor forces, ``new BR scaling" and ${\mathbf E_{sym}}$}
Another potentially important consequence of the topology change described above is on the nuclear tensor forces. It will therefore affect the EoS of compact-star matter, particularly on the symmetry energy $E_{sym}$~\cite{LPR}. This aspect, highly relevant to our main discussion, is, however, not explicitly taken into account in the paper. It will be done in a future presentation

In the effective field theory approach to nuclear forces, in which the relevant degrees of freedom are nucleons, pions and vector mesons, the nuclear tensor forces are highly sensitive to in-medium properties of the $\rho$ meson, specifically its mass and its coupling to nucleons. The $\rho$ tensor force comes with the sign opposite to that of the pion tensor, so their contributions to the tensor force tend to cancel as density increases because the strength of the $\rho$ tensor force increases at increasing density. This is because the parameters of the $\rho$ meson scale according to what is known as BR scaling~\cite{BR}, where the $\rho$ mass drops at increasing density following roughly the quark condensate in medium. There are evidences for this scaling in a variety of nuclear processes, among which the most striking one is the quenching of the Gamow-Teller matrix element responsible for the long life-time of $^{14}$C, i.e., the famous $^{14}$C dating~\cite{c14}. There may be alternative explanations for the quenching of the Gamow-Teller matrix element involving many-body potentials or many-body correlations. However, within the given effective field framework used in Ref. \cite{c14}, it provides a strong support that the BR scaling does hold up to NM density. However, there is no evidence either for or against the BR scaling of Ref. \cite{BR} above the NM density which is pertinent for neutron stars.

In \cite{LPR}, it is shown that the Skyrmion-half-Skyrmion transition can drastically modify the BR scaling~\cite{BR} at the topology-change density $\rho_{1/2}$. Skipping details, we simply state what was found in that reference. A subtle scaling change occurs at $\rho_{1/2}$ owing to the topology change which makes the $\rho$ tensor force whose magnitude increases up to the density $\rho_{1/2}$ turn over and start decreasing, ultimately get suppressed more or less completely at a density near two to three times that of NM. This topology change turns out to leave the pion tensor  unaffected. Therefore at some high density above $\rho_{1/2}$, only the pion tensor remains. In the model calculation of Ref. \cite{LPR}, the complete suppression of the $\rho$ tensor takes place at about $3\rho_0$, in the vicinity of kaon condensation. In terms of the symmetry energy, a cusp appears at $\rho_{1/2}$. Whether this is real or an artifact of the crystal structure is not clear. In any event,  this phenomenon can have a dramatic effect on processes in baryonic matter at high density, in particular on the symmetry energy in the density regime that has not been accessed by the presently available experiments. Future experiments will provide a check for this prediction.

\subsection*{(B4) Dilaton limit and suppression of the short-distance repulsion}
Consider a baryon HLS Lagrangian implemented with a dilaton scalar introduced via the trace anomaly of QCD. We call it dHLS. Such a theory at mean field is known to be able to describe NM fairly well provided the parameters of the HLS Lagrangian are endowed with suitable BR scaling~\cite{song}. The question raised is as follows: How can one use this Lagrangian to describe the chiral phase transition that is to take place at high density, say, $\rho_c$? This question was answered in \cite{SLPR}.

While it is not known how to approach systematically as function of density the chiral restoration point with this Lagrangian, it was suggested by Beane and van Kolck~\cite{beane} how to approach Weinberg's mended symmetry limit~\cite{weinberg} by taking what is called the ``dilaton limit." In that limit, the dilaton-implemented HLS Lagrangian goes over to Gell-Mann-L\'evy linear sigma model which can then be used to describe the chiral phase transition in the standard way. This limiting process involves a transformation of a chiral singlet -- which the dilaton is -- to the fourth component of chiral four-vector, namely, $\sigma$ of the sigma model. Apart from the symmetries involved, this may be caused by a sort of level crossing in terms of quark configurations. In the quark-gluon language of QCD, the dilaton could be a complicated combination of four-quark $\bar{q}^2q^2$ wavefunction and glueball wavefunction that is low-lying at low density while the $\sigma$ is the $\bar{q}q$ component which lies higher. The level crossing may occur such that at high density near the chiral restoration, it is the latter that comes down low in energy. The key point here is that as density approaches that of chiral restoration, the dilaton limit transforms dHLS to the linear sigma model endowed with the mended symmetry. It was found in Ref. \cite{SLPR} that the dilaton limit in dHLS corresponds to taking, among others,
\be
g_A=g_V\rightarrow 1\label{dilaton}
\ee
where $g_A$ is the axial coupling constant in the neutron beta decay and $g_V$ is related to the $V$-nucleon coupling as
\be
g_{VNN}=g_{\tiny hls}(1-g_V)\label{vector}
\ee
where $V=(\rho,\omega)$ are the lowest-lying vector mesons and $g_{hls}$ is the hidden gauge coupling.

There are a variety of phenomena that could follow as consequences of the approach to the dilaton limit and influence strongly high-density physics. What is closely relevant for this discussion is that as one approaches the dilaton limit (\ref{dilaton}), the $V$-nucleon coupling decreases and goes to zero at the dilaton limit. Owing to the vector manifestation of HLS~\cite{HY:PR}, the hidden gauge coupling going proportionally to the quark condensate $\la\bar{q}q\ra$ near chiral restoration will also go to zero. However the vector meson mass also goes to zero as $\sim g_{hls}$, so the ratio $g_{VNN}^2/m_V^2\sim constant$. This means that as density increases into the regime where the dilaton limit is valid, interactions that involve  vector-meson exchanges will become weak independently of the approach to the VM fixed point. In particular, the $NN$ repulsion at short distance attributed to $\omega$ exchanges will be suppressed proportionally to $(1-g_V)^2$. This would make a drastic change to the conventional thinking of the hard core repulsion that has been understood up to date to be operative at high density. Furthermore if one were to associate short-distance repulsion involving three nucleons with $\omega$ exchanges, the suppression factor would also go like $(1-g_V)^2$ and so be equally suppressed. From what density this suppression becomes active is not known, but if this mechanism is confirmed to be viable, it will certainly affect the presently available EoS' at densities beyond what has been ``measured." Among various phenomena, as noted in the Discussion section, such a suppression mechanism could invalidate the argument of Ref. \cite{pandha} -- based on short-range repulsion in nuclear interactions -- against low critical density for kaon condensation.

\newpage

\newpage
\def\bi{\bibitem}

{}

\end{document}